\documentclass{AIAA}

\usepackage{amsmath,amssymb}

\newtheorem{theorem}{Theorem}
\newtheorem{cor}{Corollary}
\newtheorem{lem}{Lemma}

\newcommand\numberthis{\addtocounter{equation}{1}\tag{\theequation}}
\newcommand\scalemath[2]{\scalebox{#1}{\mbox{\ensuremath{\displaystyle #2}}}}

\newcommand{\Z}{\mathbb{Z}}
\newcommand{\R}{\mathbb{R}}
\newcommand{\C}{\mathbb{C}}

\newcommand{\E}[1]{\mathbb{E}[{#1}]}
\newcommand{\Cov}[1]{\rm{Cov}[{#1}]}
\renewcommand{\S}{\mathbb{S}}
\newcommand{\V}[2]{\mathcal{V}_{#1}({#2})}
\newcommand{\M}[2]{\mathcal{M}_{#1}({#2})}

\newcommand{\proj}{\rm{proj}}

\renewcommand{\rm}[1]{\mathrm{#1}}
\renewcommand{\cal}[1]{\mathcal{#1}}
\newcommand{\frk}[1]{\mathfrak{#1}}

\newcommand{\tr}[1]{\rm{Tr}({#1})}
\newcommand{\diver}[1]{\rm{div}({#1})}
\newcommand{\dd}{\mathrm{d}}
\newcommand{\wh}{\widehat}

\newcommand{\ad}{\rm{ad}}
\newcommand{\adbar}{\overline{\rm{ad}}}

\newcommand{\SO}[1]{\rm{SO}_{#1}}
\newcommand{\SU}[1]{\rm{SU}_{#1}}

\newcommand{\GL}[1]{\rm{GL}_{#1}}
\newcommand{\SE}[1]{\rm{SE}_{#1}}

\newcommand{\so}[1]{\frk{so}_{#1}}

\newcommand{\se}[1]{\frk{se}_{#1}}

\newcommand{\Aut}[1]{\rm{Aut}({#1})}
\newcommand{\Lie}[1]{\rm{Lie}({#1})}
\newcommand{\BCH}[2]{\rm{BCH}({#1},{#2})}

\begin{document}

\title{Dynamics on Lie groups with applications to attitude estimation}

\author{T.~Forrest~Kieffer\footnote{Mathematician,~Weapon~Control~Concepts~Development~Group,~Air~and~Missile~Defense~Sector;~Thomas.Kieffer@jhuapl.edu~(Corresponding Author)}~and~Michael~L.~Wall\footnote{Chief~Scientist,~Weapon~Control~Concepts~Development~Group,~Air~and~Missile~Defense~Sector}\footnote{Alphabetical author order; both authors contributed equally to this work}}\affiliation{Johns Hopkins University Applied Physics Laboratory, Laurel, MD 20723}

\begin{abstract}
The problem of filtering -- propagation of states through stochastic differential equations (SDEs) and association of measurement data using Bayesian inference -- in a state space which forms a Lie group is considered. Particular emphasis is given to concentrated Gaussians (CGs) as a parametric family of probability distributions to capture the uncertainty associated with an estimated state. The so-called group-affine property of the state evolution is shown to be necessary and sufficient for linearity of the dynamics on the associated Lie algebra, in turn implying CGs are invariant under such evolution. A putative SDE on the group is then reformulated as an SDE on the associated Lie algebra. The vector space structure of the Lie algebra together with the notion of a CG enables the leveraging of techniques from conventional Gaussian-based Kalman filtering in an approach called the \textit{tangent space filter} (TSF). We provide example calculations for several Lie groups that arise in the problem of estimating position, velocity, and orientation of a rigid body from a noisy, potentially biased inertial measurement unit (IMU). For the specific problem of attitude estimation, numerical experiments demonstrate that TSF-based approaches are more accurate and robust than another widely used attitude filtering technique.
\end{abstract}

\maketitle

%
%

\section*{Nomenclature}

\noindent\begin{tabular}{@{}lcl@{}}
$\Z$ &=& $\{ 0 , \pm 1 , \pm 2 , \cdots \}$, the set of integers.\\
$\R$ &=& the set of real numbers.\\
$\R^n$ &=& the set of $n$-dimensional vectors with real entries.\\
$\R^{n \times m}$ &=& the set of $n \times m$ matrices with real entries.\\
$I_n$ &=&  the $n \times n$ identity matrix.\\
$\langle \cdot , \cdot \rangle$ &=& standard Euclidean inner product on $\R^n$.\\
$[\cdot]_{\times}$ &=& cross product matrix satisfying $[a]_{\times} b = a \times b$ for $a,b \in \R^3$.\\
$[ A , B ]$ &=& $AB - BA$, the commutator of $A,B \in \R^{n \times n}$.\\
$\S^3$ &=& $\{ q \in \R^4 : \| q \| = 1 \}$, $3$-dimensional sphere.\\
$\GL{n}$ &=& $\{ A \in \R^{n \times n} : \det{A} \neq 0 \}$, the general linear group in $n$ dimensions.\\
$\SO{n}$ &=& $\{ A \in \R^{n \times n} : A A^T = A^T A = I_n \}$, the special orthogonal group in $n$ dimensions.\\
$\so{n}$ &=& $\{ A \in \R^{n \times n} : A^T = - A \}$, the Lie algebra of $\SO{n}$.\\
$\SE{n}$ &=& $\SO{n} \ltimes \R^n$, the special Euclidean group in $n$ dimensions.\\
$\se{n}$ &=& the Lie algebra of $\SE{n}$.\\
$G$ &=& matrix Lie group (i.e., a closed subgroup of $\GL{d}$ for some $d > 0$).\\
$\frk{g}$ &=& the Lie algebra associated with $G$, often denoted $\Lie{G}$.\\
$\cal{V}_{\frk{g}}$ &=& "vectorization" map associating the Lie algebra $\frk{g}$ with $\R^{\dim{\frk{g}}}$ as $X \in \frk{g} \mapsto \V{\frk{g}}{X} \in \R^{\dim{\frk{g}}}$.\\
$\cal{M}_{\frk{g}}$ &=& "matrization" map associating $\R^{\dim{\frk{g}}}$ with $\frk{g}$, i.e. the inverse of $\cal{V}_{\frk{g}}$.\\
$\ad_X$ &=& the adjoint action on $\frk{g}$ given by $\ad_X Y = [X , Y]$ with $X , Y \in \frk{g}$.\\ 
$J_{\frk{g}}(X)$ &=& $\int_0^1 e^{- s \ad_X } \dd s$, for $X \in \frk{g}$, is the Jacobian of the exponential map.\\
$W_{\!\frk{g}} (X)$ &=& $J^{-1}_{\frk{g}} (-X)$, for $X \in \frk{g}$.\\
$\adbar_{\xi_1} \xi_2$ &=& $\V{\frk{g}}{[\M{\frk{g}}{\xi_1} , \M{\frk{g}}{\xi_2}]}$ for $\xi_1 , \xi_2 \in \R^{\dim{\frk{g}}}$.\\
$\overline{J}_{\!\frk{g}}(\xi)$ &=& $\int_0^1 e^{- s \adbar_{\xi} } \dd s$, for $\xi \in \R^{\dim{\frk{g}}}$.\\
$\overline{W}_{\!\!\frk{g}} (\xi)$ &=& $\overline{J}^{-1}_{\frk{g}} (-\xi)$, for $\xi \in \R^{\dim{\frk{g}}}$.\\
$\BCH{X_1}{X_2}$ &=& $\log{(e^{X_1} e^{X_2})}$ for $X_1, X_2\in \frk{g}$.\\
$\cal{T}_G ( \mu , \Sigma )$ &=& CG distribution on $G$ with group mean $\mu \in G$ and covariance matrix $\Sigma$.\\
$\cal{N}^n (\mu , \Sigma)$ &=& $n$-dimensional Gaussian distribution with mean $\mu \in \R^n$ and covariance $\Sigma \in \R^{n \times n}$
\end{tabular}

\clearpage

\section{Introduction}

The impact of the Kalman filter (KF) in statistical estimation and control theory is difficult to overstate. The KF enables the calculation of the joint probability distribution of a vector of {state} variables $\mathbf{x}$ from a collection of noisy measurements supplemented with linear models of the state vector dynamics and measurement~\cite{kalman1960new,kalman1961new}. The key assumptions required for the KF to be optimal -- that dynamics and measurements are linear in the state space variables, that the noise appearing in dynamics and measurements is additive and white, and that the state vector distribution is initially Gaussian -- are physically reasonable or can be made to hold approximately with good accuracy in many cases. This has led to the KF being used ubiquitously in guidance, navigation, and control~\cite{grewal2007global}, in robotics, in space and defense~\cite{bar2004estimation,ristic2004beyond}, and myriad other applications. In places where the linearity assumption breaks down, the extended Kalman filter (EKF), in which the nonlinear models are linearized about the current state estimate point, or the unscented Kalman filter (UKF), in which a collection of nonlinear model outcomes are utilized to estimate state vector properties~\cite{julier1997new}, are widely employed, often to good accuracy. Some exact results are known for distributions other than Gaussians, though these usually require specific dynamical and measurement models to ensure the that resulting filter process leaves the form of the non-Gaussian distribution invariant and only updates its parameters; the work of Bene\u{s}~\cite{benevs1981exact} and Daum~\cite{daum2005nonlinear} is paradigmatic of this approach. At the extreme of nonlinear filtering is particle filters~\cite{gordon1993novel}, which represent the filtered distribution as a weighted collection of delta functions~\cite{ristic2004beyond} and so place no constraints on dynamical or measurement models or the underlying state vector distribution. However, particle filters are limited by a sampling error which can be slow to converge, especially for large state space dimension~\cite{daum2003curse}, and this can pose challenges for stability and efficient implementation.


A more subtle setting for nonlinear filtering is estimation of a state which is subject to a nonlinear constraint. Here, several complications can occur: linearization of the constraint may be inaccurate across the support of the state vector distribution and so key properties are not maintained. The constraint also reduces the effective dimensionality of the state vector, which introduces singularities into its associated covariance matrix that are known to cause issues for filtering. In such cases, a modification of the distribution is required to avoid non-physical predictions, but such modification will likely be at odds with the Gaussian assumptions required for the standard KF. 

The present work investigates a special case of this situation in which the constraint is formulated as a requirement that the state vector lives on a Lie group, which is a mathematical space that is both a smooth manifold and a group. In this setting, the filtering problem may be translated from the Lie group to its associated tangent spaces (i.e., its Lie algebra), which enables the constraint to be maintained globally through the exponential map connecting the tangent spaces with the Lie group, allows for accurate linearization due to the "flat nature" of the tangent spaces, and enables an unconstrained, singularity-free representation of the state and covariance.


Before presenting our general theoretical results, we introduce the motivating application of attitude estimation from noisy, potentially biased gyro measurements in Sec.~\ref{sec:AttEst}. In addition to providing motivation for our research, this also allows for concrete example calculations alongside the theory. Our theoretical exposition starts with the most general setting in Section \ref{sec:Characterization_of_group-affine_vector_fields}, which is that of an arbitrary matrix Lie group $G$ with Lie algebra $\frk{g}$ endowed with an evolution equation of the form $\dot{g} = f(g)$ for some vector field $f : G \rightarrow \frk{g}$. The question of the necessary and sufficient conditions on $f$ for which the corresponding dynamics on $\frk{g}$ are linear is then addressed. Leveraging the results of \cite{barrau2019linear}, $f$ must satisfy the so-called \emph{group-affine property} and we go beyond prior work to give a complete characterization of group-affine vector fields (see Theorem \ref{thm:Characterization}).  

Section \ref{sec:CG} formally introduces the class of CGs on a matrix Lie group. The main result here is Corollary \ref{cor:1}, which makes precise the claim that the class of CGs are invariant under the flow generated by a group-affine vector field. This Corollary provides the theoretical foundation for one of the main contentions in this work, namely that CGs are the most natural choice of probability distributions to filter with on a Lie group. Particular emphasis is then given to choosing a group law on the underlying state space which makes the state space dynamics in question group-affine. Section \ref{sec:TSSDE} considers an application-motivated stochastic generalization of group-affine dynamics and derives the corresponding SDE on $\frk{g}$. This SDE on $\frk{g}$ will be referred to as the \emph{tangent space SDE} (TSSDE) on $\frk{g}$ and is, in general, an SDE with state-dependent diffusion. This section ends with a discussion regarding the invariance of CGs in the presence of noise. 

The theoretical results generated in Section \ref{sec:Characterization_of_group-affine_vector_fields}, \ref{sec:CG}, and \ref{sec:TSSDE} are used as motivation for our Lie group filtering methodology, detailed in Section \ref{sec:Filtering_on_Lie_groups}. After introducing a discrete-time observation process, we design an (approximate) nonlinear filtering algorithm, called the TSF, that is based on the unscented transform (UT). Key novelties of this algorithm are the use of the continuous-time unscented transform (CTUT) \cite{sarkka2007unscented} (generalized to state-dependent diffusion) to propagate moments through the TSSDE and a \textit{whitening} procedure based on the Baker–Campbell–Hausdorff (BCH) formula. We also suggest how the TSF may be made more computationally efficient by considering a short-time expansion of the Fokker-Planck equation (FPE) associated with the TSSDE and deriving moment evolution equations based off this to replace to CTUT for moment propagation.

In Section \ref{sec:Examples}, three application-relevant examples are worked out in detail. These three examples correspond to the well-known group $\SU{2}$, the special Euclidean group $\SE{3}$, and the group $\SE{2,3}$, whose underlying state space is $\SO{3} \times \R^6$ and is a simple generalization of $\SE{3}$. The $\SU{2}$ example is motivated by the problem of attitude filtering under the paradigm of dynamical model replacement (DMR) without gyro bias, while the $\SE{2,3}$ example is motivated by the inertial navigation system (INS) state estimation problem without IMU biases. The group $\SE{3}$ provides a bridge between these two examples. The natural dynamics for the $\SU{2}$ problem falls into the class of constant Lie group dynamics, which are the simplest to analyze. On the other hand, the natural dynamics motivated by the INS problem are not constant, but fall into the more general class of group-affine dynamics. Emphasis is placed on how the $\SE{2,3}$ group law on the underlying state space $\SO{3} \times \R^6$ is the choice which results in group-affine dynamics (as compared to, e.g., the direct product (DP) group law). Additional insights regarding the fundamental solution of the FPE associated with the $\SU{2}$ problem are provided at the end of Section \ref{sub:SU2}.

A fourth example, covered in Section \ref{sec:GyroBias}, is the Lie group whose underlying manifold is $\SO{3} \times \R^3$. This group is motivated by the problem of attitude estimation with the inclusion of the gyro bias that is to-be-estimated. This example will highlight the interplay between the equations of motion (i.e., dynamics), which are motivated by physics, and the freedom to choose a group structure which in turn makes the associated vector field group-affine. In this fourth example, we argue that there is no \textit{fixed} group structure on $\SO{3} \times \R^3$ with makes the attitude dynamics with gyro bias group-affine. Instead, a time-parameterized family of group laws must be considered to yield a vector field which is \textit{almost} group-affine. This time-parameterized family of group laws is approximately $\SE{3}$ to zeroth-order in time. (This observation is consistent with the proposal to use the $\SE{3}$ group law for treating gyro bias in, e.g., \cite{barrau2022geometry}.) This section ends with a numerical experiment that compares performance of the TSF using two different group laws on $\SO{3} \times \R^3$, namely that induced by the DP and that of a semidirect product (i.e., $\SE{3}$), to the Unscented Quaternion Estimator (USQUE) \cite{Crassidis_UKF}. It is shown that the TSF using the $\SE{3}$ group law exhibits near-perfect filter consistency, far outperforming the USQUE and TSF using the DP group law (Figure~\ref{fig:ChiSq}).

\section{Attitude descriptions and estimation}\label{sec:AttEst}

In attitude filtering, one is concerned with estimating an element of the non-abelian Lie group of $3$-dimensional spatial rotations, denoted as $\SO{3} := \{ R \in \R^{3 \times 3} : R R^T = I_3 , ~ \det{R} = 1\} $. Specifically, a matrix $R \in \SO{3}$ describes a rotation from one reference frame (e.g., the body frame of a rigid body) to another reference frame (e.g., an inertial or Earth-fixed frame). Since $\SO{3}$ is a $3$-dimensional manifold, one may  use, at least locally, a set of $3$ real parameters (e.g., Euler angles) to describe a matrix $R \in \SO{3}$. However, since $\SO{3}$ is not homeomorphic to $\R^3$, a single set of such parameters will necessarily have a singularity at some point (e.g., Euler angles have a singularity that cause the problem of gimbal lock). 

The standard approach to avoid the issue of coordinate singularities is to work with a higher dimensional parameterization of rotations, such as the unit quaternions. A quaternion may be thought of as a vector $q \in \R^4$ which is partitioned as $q = ( v ~  s )$, where $v = ( v_1 ~ v_2 ~ v_3 ) \in \R^3$ is the "vector" part and $s \in \R$ is the "scalar" part. A unit quaternion $q$ has $\|q\| = 1$, where $\| \cdot \|$ indicates the standard Euclidean length in $4$ dimensions. The set of all unit quaternions is the $3$-sphere, denoted as $\S^3 := \{ q \in \R^4 : \| q \| = 1 \}$. Given a unit quaternion $q \in \S^3$, there is a map $R : \S^3 \rightarrow \SO{3}$ that associates the unit quaternion $q$ with a rotation $R(q)$ between two specified reference frames. Explicitly,
\begin{align}\label{eq:Rorth}
R(q) = I_3 - 2 s [v]_{\times} + 2 [v]_{\times}^2 ,
\end{align}
where $[v]_{\times} \in \R^{3 \times 3}$ is the \textit{cross product matrix}, which is the matrix satisfying $[v]_{\times} w = v \times w$ for all $w \in \R^3$. Note that $R (q) R^T (q) = I_3$ and $\det{R(q)} = 1$ is a result of the constraint $\|q\| = 1$. 

One may endow $\S^3$ with a multiplication operation $\otimes$ such that the multiplication of two unit quaternions describes the composition of their individual rotations, i.e. $R (q \otimes q') = R(q) R(q')$. (Note that we are using M. Shuster's convention for quaternion multiplication \cite{shuster1993survey}. This choice is made so that the map $R$ in \eqref{eq:Rorth} is a (Lie) group homomorphism $\S^3 \rightarrow \SO{3}$. The reader should be aware that another convention is in use in the literature and which convention is being used should always be verified for consistency.) It may be shown that $q \otimes q'$ defines another unit quaternion given by
\begin{align}\label{eq:QuatMult}
q \otimes q'  = \left( \begin{array}{c} s v' + s' v - v \times v' \\ s s' - \langle v , v' \rangle \end{array} \right) ,
\end{align}
where $\langle v , v' \rangle = v_1 v_1' + v_2 v_2' + v_3 v_3'$ indicates the standard Euclidean inner product of the vectors $v , v' \in \R^3$. This multiplication operation can be represented in matrix form as 
\begin{align}\label{eq:Quant_as_matrix}
\left[q \otimes\right]  : = s I_4 + \M{\frk{s}}{v} , 
\end{align}
where $\cal{M}_{\frk{s}} : \R^3 \rightarrow \R^{4 \times 4}$ is the anti-symmetric matrix
\begin{align}\label{def:Ms}
\M{\frk{s}}{v} = \left(\begin{array}{cc} - [v]_{\times} & v \\[2pt] 
- v^T & 0 \end{array}\right) ,
\end{align}
in which $v$ is interpreted as a column vector and $v^T$ as a row vector. Under the operation $\otimes$ the inverse quaternion is given by $q^{-1} = ( - v ~ s )$ and the identity quaternion is given by $(0 ~ 1)$. Hence, $\S^3$ endowed with quaternionic multiplication forms a Lie group. As a Lie group, $\S^3$ is recognized as the double cover of $\SO{3}$. The term \textit{double cover} refers to the fact that the map \eqref{eq:Rorth} satisfies $R(q) = R(-q)$. In other words, there are two distinct unit quaternions that yield the same $\SO{3}$-matrix $R$. Finally, it is worth noting that the Lie group $\S^3$ is typically identified with the special unitary group $\SU{2} = \{ U \in \C^{2 \times 2} : U U^{\dagger} = I_3 , ~ \det{U} = 1 \} $.

If a rigid body is subject to an angular velocity $\omega (t) \in \R^3$ relative to a chosen reference frame and $R (t) \in \SO{3}$ describes its orientation relative to that frame, then $\dot{R} = [\omega]_{\times} R$, where the "dot" denotes the time derivative. This equation of motion for  rotations may be recast in terms of quaternions using the inverse of the map \eqref{eq:Rorth}, which results in
\begin{align}\label{eq:qEOM}
\left\lbrace \begin{array}{l}
\dot{q} = \dfrac{1}{2} \M{\frk{s}}{\omega} q \\[2pt]
q(0) = q_0 \in \S^3 ,
\end{array} \right.
\end{align}
where $q_0$ describes the initial orientation of the rigid body. In the case of deterministic and constant $\omega (t) = \omega_0$, the solution is given by the matrix exponential:
\begin{align}\label{eq:qEOM_soln}
q(t) = e^{t \M{\frk{s}}{\omega_0} / 2} q_0 . 
\end{align}

Typically, the angular rate $\omega$ that is passed to \eqref{eq:qEOM} is a noisy measurement obtained from a gyroscope. A common parameterization of angular rate measurements is given by the \textit{Farrenkopf model} \cite{farrenkopf1978analytic, Crassidis_Extension}. Stated as a system of coupled SDEs, the Farrenkopf model reads
\begin{align}\label{eq:Farrenkopf}
\left\lbrace \begin{array}{l}
\omega_m \dd t = \left( \omega + \beta \right) \dd t + \dd \eta \\[2pt]
\dd \beta = \dd\zeta ,
\end{array} \right.
\end{align}
in which $\omega_m$ is the measured rate, $\omega$ is the true rate, $\beta$ is the rate bias, and $\dd\eta$ and $\dd\zeta$ are vectors of mutually uncorrelated Wiener processes whose derivatives are delta-correlated white noises with spectral density matrices $Q_{\eta}$ and $Q_{\zeta}$, respectively. This model excludes gyro misalignment and scale factor errors, but they can be straightforwardly included. 

It must be stressed that the vector $\omega_m$ is a measurement, and so, strictly speaking, should be estimated when utilized in a filter~\cite{Crassidis_Extension}. Instead, the conventional approach, known as \emph{dynamic model replacement} (DMR)~\cite{Crassidis_UKF}, is followed. DMR uses the measurements $\omega_m$ directly in the model and treats this measured angular rate as a constant over the propagation time step. The SDE describing the evolution of the quaternion under the Farrenkopf model is then determined by solving eqn.~\eqref{eq:Farrenkopf} for $\omega$ and plugging the result into eqn.~\eqref{eq:qEOM}. The result is the \textit{quaternion SDE}:
\begin{align}\label{eq:QSDE}
\left\lbrace \begin{array}{l}
\dot{q} = \dfrac{1}{2} \M{\frk{s}}{\omega_m - \beta - \eta} q  \\[2pt]
\dot{\beta} = \zeta \\[1pt]
\| q \| = 1 .
\end{array} \right.
\end{align}
As an SDE, \eqref{eq:QSDE} must carry the Stratonovich interpretation, otherwise solutions may not maintain the constraint $\|q\| = 1$. Eqn.~\eqref{eq:QSDE} represents the SDE relevant for propagation in an attitude filter based on DMR using the Farrenkopf model \eqref{eq:Farrenkopf}. 

A key question not yet addressed is: what is an appropriate family of probability distributions to describe the uncertainty associated with $q$? In this regard, let $\delta \sim \cal{N}^3 (0 , \Sigma)$ and $\mu \in \S^3$ be some fixed reference quaternion. Consider the distribution over $\S^3$ induced by $q = e^{\M{\frk{s}}{\delta}} \mu$ (c.f. \eqref{eq:qEOM_soln}). The distribution associated with the random quaternion $q$ is an example of a CG on the Lie group $\S^3$ with parameters $\Sigma$ and $\mu$. With this family of distributions identified, the next question of immediate practical importance is: how do the parameters of the distribution evolve under \eqref{eq:QSDE}? Here, we find our first connection with the main results of our work. Namely, if $q_0 = e^{\M{\frk{s}}{\delta_0}} \mu_0$ for some $\delta_0 \sim \cal{N}^3 (0,\Sigma_0)$ and $\mu_0 \in \S^3$, then the solution \eqref{eq:qEOM_soln} is of the form $q (t) = e^{\M{\frk{s}}{\delta (t)}} \mu (t)$ for some $\delta (t) \sim \cal{N}^3 (0,\Sigma (t))$ and $\mu (t) \in \S^3$. In other words, if $q_0$ is CG distributed, then $q (t)$ in \eqref{eq:qEOM_soln} is also CG distributed (c.f.~Corollary \ref{cor:1}). The remainder of this section is concerned with explicitly demonstrating this claim. First, it is necessary to collect a few more facts about $\S^3$ and its Lie algebra.

Eqn.~\eqref{eq:Quant_as_matrix} gives us a convenient faithful representation $\rho$ of the group $\S^3$ as a subgroup of $\SO{4}$, which is the Lie group of $4$ dimensional rotations, a higher dimensional generalization of $\SO{3}$. (In what follows, we will write $\rho (\S^3)$ to denote the subgroup of $\SO{4}$ representing unit quaternions.) The convenience comes, in part, from the determination of the Lie algebra of $\rho (\S^3)$, namely the set
$\frk{s} := \{ \M{\frk{s}}{a} : a \in \R^3 \} \subset \so{4} $
where $\so{4}$ is the set of all $4 \times 4$ antisymmetric matrices (i.e., the Lie algebra of $\SO{4}$). That $\frk{s}$ is the Lie algebra of $\rho (\S^3)$ follows from the identity
\begin{align}\label{eq:LieGroupExp_S3}
e^{\M{\frk{s}}{a}} = \cos{( \| a \| )} I_4 + \frac{\sin{(\| a \|)}}{\|a\|} \M{\frk{s}}{a} \in \rho (\S^3) ,
\end{align} 
which follows from the Taylor series expansion of the matrix exponential and $\M{\frk{s}}{a}^2 = - \| a \|^2 I_4$. 

Formula \eqref{eq:LieGroupExp_S3} is an instance of the \textit{exponential map} $\exp : \frk{g} \rightarrow G$ from the Lie algebra $\frk{g}$ of a Lie group $G$, which is a concept that will be dealt with regularly in the sequel. The exponential map is defined for abstract Lie groups and their associated Lie algebras, but for the case of matrix Lie groups (i.e., closed subgroups of $\GL{n}$), the abstract exponential map coincides with the standard matrix exponential. The Lie bracket on $\{ \M{\frk{s}}{a} : a \in \R^3 \}$ is given by
\begin{align}\label{eq:Omega_Commutator}
[ \M{\frk{s}}{a} , \M{\frk{s}}{b} ] := \M{\frk{s}}{a}\M{\frk{s}}{b} - \M{\frk{s}}{b}\M{\frk{s}}{a} = - 2 \M{\frk{s}}{a\times b} .
\end{align}
In other words, the map $a \mapsto \M{\frk{s}}{a}$ is a Lie algebra isomorphism from the Lie algebra $(\R^3 , \times)$ to the matrix Lie algebra $\frk{s}$. The physical interpretation of $e^{\M{\frk{s}}{a}/2}$ is that, when applied to a unit quaternion, it yields a rotation through the angle $\|a\|$ about the axis $a / \|a\|$. 

Let $\omega_0 \in \R^3$. Suppose $q(t) = e^{t \M{\frk{s}}{\omega_0} / 2} q_0$ where $q_0 = e^{\M{\frk{s}}{\delta_0}} \mu_0$ with $\delta_0 \sim \cal{N}^3 (0,\Sigma_0)$ and $\mu_0 \in \S^3$. The goal is to show that $q (t)$ is still of the form $q (t) = e^{\M{\frk{s}}{\delta (t)}} \mu (t)$ for some $\delta (t) \sim \cal{N}^3 (0 , \Sigma (t))$ and $\mu (t) \in \S^3$. The first step is to observe that
\begin{align*}
q(t) = e^{t \M{\frk{s}}{\omega_0} / 2} e^{\M{\frk{s}}{\delta_0}} \mu_0 = e^{t \M{\frk{s}}{\omega_0} / 2} e^{\M{\frk{s}}{\delta_0}} e^{-t \M{\frk{s}}{\omega_0} / 2} \mu (t) ,
\end{align*}
where $\mu (t) = e^{t \M{\frk{s}}{\omega_0} / 2} \mu_0$. What remains is to demonstrate that $e^{t \M{\frk{s}}{\omega_0} / 2} e^{\M{\frk{s}}{\delta_0}} e^{-t \M{\frk{s}}{\omega_0} / 2} = e^{\M{\frk{s}}{\delta (t)}}$ for some $\delta (t) \in \R^3$.

Recall a standard result in Lie theory, namely the \textit{braiding identity} \cite[Proposition 3.35]{hall2013lie}: 
\begin{align}\label{eq:braiding}
e^X Y e^{-X} = e^{\ad_X} Y ,
\end{align}
which holds for all $X , Y \in \frk{g}$. In \eqref{eq:braiding}, $\ad_X : \frk{g} \rightarrow \frk{g}$ is given by $\ad_X Y := [X,Y]$ and denotes the \textit{adjoint action} on the Lie algebra $\frk{g}$. Let $a , b \in \R^3$. The braiding identity \eqref{eq:braiding}, the commutator identity \eqref{eq:Omega_Commutator}, and linearity of $\cal{M}_{\frk{s}}$, yields
\begin{align*}
e^{\M{\frk{s}}{a}} e^{\M{\frk{s}}{b}} e^{-\M{\frk{s}}{a}} = e^{e^{\M{\frk{s}}{a}} \M{\frk{s}}{b} e^{-\M{\frk{s}}{a}}} = e^{ e^{\ad_{\M{\frk{s}}{a}}} \M{\frk{s}}{b} } = e^{ \M{\frk{s}}{ e^{ - 2 [a]_{\times} } b } } . 
\end{align*}
Applying the previous to identity to the expression $e^{t \M{\frk{s}}{\omega_0} / 2} e^{\M{\frk{s}}{\delta_0}} e^{-t \M{\frk{s}}{\omega_0} / 2}$, one finds that the desired $\delta (t)$ is given by $\delta (t) = e^{- t [\omega_0]_{\times} } \delta_0$. In other words, $\delta (t) \sim \cal{N}^3 (0 , \Sigma (t))$ where $\Sigma (t) = e^{- t [\omega_0]_{\times} } \Sigma_0 e^{t [\omega_0]_{\times} }$. What has just been demonstrated is the invariance of the class of CGs on $\S^3$ under the flow generated by the dynamics \eqref{eq:qEOM}. The deeper theory that is developed in Section \ref{sec:Characterization_of_group-affine_vector_fields} informs us that this invariance is really due to the fact that the vector field $q \mapsto \frac{1}{2} \M{\frk{s}}{\omega} q$ is \textit{group-affine} on $\S^3$. 

In general, additional uncertainty contributes to the attitude quaternion during propagation due to bias and noise corrupting the measured angular rates, as in \eqref{eq:Farrenkopf}. Some natural questions are then: are CGs invariant under propagation in the presence of bias and noise? If not, how should a CG approximation be maintained during propagation? These and related questions are addressed in generality throughout the remainder of this manuscript. In particular, in Section \ref{sec:Filtering_on_Lie_groups}, the main result of Section \ref{sec:TSSDE} is used together with the CTUT to (approximately) propagate the parameters of a CG forward in time in the presence of noise. Once these tools are in hand, the problem of attitude filtering (with the inclusion of gyro bias) is revisited in Section \ref{sec:GyroBias}. 

\section{Characterization of group-affine vector fields}\label{sec:Characterization_of_group-affine_vector_fields}

The main theoretical exposition begins with a determination on necessary and sufficient conditions for dynamics on a Lie group to result in linear dynamics on the associated Lie algebra. Let $G$ be a matrix Lie group of dimension $n$, and $\frk{g} = \Lie{G}$ be its Lie algebra. Throughout this manuscript, $G$ is assumed to have surjective exponential map. Consider an initial value problem (IVP) on $G$ of the form
\begin{align}\label{eq:generalIVP}
\left\lbrace \begin{array}{l}
\dot{g} = f (g) \\
g(0) = g_0 \in G ,
\end{array} \right. 
\end{align}
where $f : G \rightarrow \frk{g}$ is a smooth map and the dot indicates time derivative. Here, $f$ is taken to be time-independent to avoid technical complications but this isn't necessary. Suppose that the solution $g : [0,\infty) \rightarrow G$ to \eqref{eq:generalIVP} satisfies $g (t) = e^{X(t)} \mu (t)$, where $X (t)$ has $X (0) = X_0 \in \frk{g}$, and $\mu (t)$ satisfies an ordinary differential equation (ODE) analogous to the first equation in \eqref{eq:generalIVP}:
\begin{align*}
\left\lbrace \begin{array}{l}
\dot{\mu} = f (\mu) \\
\mu(0) = \mu_0 \in G ,
\end{array} \right.
\end{align*}
with $\mu_0 \in G$ known a priori. Our goal is to find the corresponding ODE for $X (t) \in \frk{g}$, and then determine the conditions on $f$ that are required to make this ODE for $X$ \textit{affine} (in the sense of affine maps on $\R^n$). 

In order to derive an ODE for $X$, we recall the result in Lie theory known as the \textit{derivative of the exponential map} \cite[Theorem 5.4]{hall2013lie}. Let $X : \R \rightarrow \frk{g}$ be a $C^1$-path in the Lie algebra. The derivative of the exponential map is given by
\begin{align}\label{eq:DerivativeExpMap}
\frac{\dd}{\dd t} e^{X (t)} = e^{X (t)} J_{\frk{g}} (X (t)) \dot{X} (t) ,
\end{align}
where $J_{\frk{g}}(X)$ is referred to as the \textit{Jacobian of the exponential map} and is given by
\begin{align}\label{eq:JacobianExpMap}
J_{\frk{g}}(X) = \int_0^1 e^{- s \ad_X } \dd s .
\end{align}
(Note that some authors define the Jacobian with the $e^X$ on the right hand side (RHS) of \eqref{eq:DerivativeExpMap} included.) Taking the time derivative of $g (t) = e^{X(t)} \mu (t)$, while using \eqref{eq:DerivativeExpMap} and \eqref{eq:generalIVP}, results in $e^X J_{\frk{g}} (X) \dot{X} \mu + e^X f (\mu) = f (e^X \mu)$, or, equivalently,
\begin{align}\label{eq:ODE_xi}
\left\lbrace \begin{array}{l}
\dot{X} = J_{\frk{g}}^{-1} (X) ( e^{-X} f (e^X \mu) - f (\mu) ) \mu^{-1} \\[3pt]
X(0) = X_0 \in \frk{g} .
\end{array} \right. 
\end{align}

The goal now is to determine a condition on $f$ that makes the RHS of \eqref{eq:ODE_xi} a linear (affine) map of $X$. The relevant condition here is referred to as the \textit{group-affine property} \cite{barrau2019linear}. A vector field $f : G \rightarrow \frk{g}$ is said to be \textit{group-affine} if, for all $g_1 , g_2 \in G$,
\begin{align}\label{eq:GroupAffineCondition}
f (g_1 g_2) = f (g_1) g_2 + g_1 f (g_2) - g_1 f (I) g_2 ,
\end{align} 
where $I \in G$ is the identity element. An important class of examples of group-affine vector fields are those of the form $f(g) = X_1 g + g X_2$ for $X_1 , X_2 \in \frk{g}$. The ODE in \eqref{eq:generalIVP} with $f(g) = X_1 g$ is referred to as the (left) Poisson equation on the group $G$ in some of the literature (see, e.g., \cite{muller2021review}). However, in this manuscript, this special case is referred to as "\textit{constant}" dynamics on the group, using an analogy with $\R^n$. Constant vector fields will be analyzed in detail in Section \ref{sec:TSSDE}. As shown in Theorem \ref{thm:Characterization}, not all group-affine vector fields are constant.

It is not entirely trivial to see how \eqref{eq:GroupAffineCondition} forces \eqref{eq:ODE_xi} to simplify to a linear equation. Indeed, if $f$ is assumed to be group-affine and the identity \eqref{eq:GroupAffineCondition} is applied to simplify the RHS of \eqref{eq:ODE_xi}, then
\begin{align}\label{eq:ODE_xi_groupaffine}
\left\lbrace \begin{array}{l}
\dot{X} = J_{\frk{g}}^{-1} (X) (e^{-X} f (e^X) - f (I)) \\[3pt]
X(0) = X_0 \in \frk{g} .
\end{array} \right. 
\end{align}
Strikingly, while the dependence on $\mu$ was eliminated from the RHS of \eqref{eq:ODE_xi}, eqn.~\eqref{eq:ODE_xi_groupaffine} does not appear to reduce to a linear equation in $X$. Additional facts regarding group-affine vector fields beyond condition \eqref{eq:GroupAffineCondition} are needed to argue that \eqref{eq:ODE_xi_groupaffine} is indeed a linear system. 

Let $\psi_t : G \rightarrow G$ denote the flow associated with the equation $\dot{g} = f(g)$, where $f$ satisfies \eqref{eq:GroupAffineCondition}. The reason why \eqref{eq:GroupAffineCondition} is called group-affine is owed to \cite[Proposition 8]{barrau2019linear}, which asserts that, for each $t > 0$, there exists an element $\varphi_t \in \Aut{G}$ of the automorphism group $\Aut{G}$ of $G$ and $\nu_t \in G$ such that
\begin{align}\label{eq:Flow_of_GroupAffine_VectorField}
\psi_t (g_0) = \varphi_t (g_0) \nu_t  .
\end{align}
The terminology of "group-affine" now becomes clear by analogy with linear systems on $(\R^n , +)$ since automorphisms of $(\R^n , +)$ are precisely linear maps. Eqn.~\eqref{eq:Flow_of_GroupAffine_VectorField} is key to proving one implication in the following Theorem. Before presenting this Theorem, recall that a derivation is a linear map $D : \frk{g} \rightarrow \frk{g}$ that satisfies $D [X,Y] = [DX,Y] + [X,DY]$.

\begin{theorem}\label{thm:Characterization}
A vector field $f : G \rightarrow \frk{g}$ is group-affine \eqref{eq:GroupAffineCondition} if and only if there exists a derivation $D : \frk{g} \rightarrow \frk{g}$ and $Y_1 , Y_2 \in \frk{g}$ such that, for all $X \in \frk{g}$,
\begin{align}\label{eq:GroupAffineCharacterization}
f(e^X) = e^X J_{\frk{g}}(X) D X + e^X Y_1  + Y_2 e^X ,
\end{align}
where $J_{\frk{g}}(X)$ is given by \eqref{eq:JacobianExpMap}. As a consequence, for a group-affine vector field, the ODE \eqref{eq:ODE_xi_groupaffine} simplifies to
\begin{align}\label{eq:ODE_xi_groupaffine_thm}
\left\lbrace \begin{array}{l}
\dot{X} = ( D + \ad_{Y_2} ) X \\[3pt]
X(0) = X_0 \in \frk{g} .
\end{array} \right. 
\end{align}
The appearance of $Y_2$ in \eqref{eq:ODE_xi_groupaffine_thm} instead of $Y_1$ is a consequence of our choice to represent $g \in G$ as $e^X \mu$ instead of $\mu e^X$.
\end{theorem}

\textit{Proof.} It is first argued that, if $f : G \rightarrow \frk{g}$ is group-affine, then there exists a derivation $D : \frk{g} \rightarrow \frk{g}$ such that $f$ has the form \eqref{eq:GroupAffineCharacterization}. This argument will leverage three keys facts: 1) the flow associated with $f$ is of the form \eqref{eq:Flow_of_GroupAffine_VectorField}, 2) an automorphism of a Lie group induces an automorphism of the Lie algebra \eqref{eq:Thm56Hall}, and 3) the Lie algebra of $\Aut{\frk{g}}$ (viewed as a Lie group) is equal to the set of derivations on $\frk{g}$. 

The starting point to show the RHS of \eqref{eq:ODE_xi_groupaffine} is a linear function of $X$ is \cite[Theorem 5.6]{hall2013lie}, which asserts the existence of a unique $T_t \in \Aut{\frk{g}}$ such that
\begin{align}\label{eq:Thm56Hall}
\varphi_t (e^X) = e^{T_t (X)} , \hspace{5mm} \forall X \in \frk{g} ,
\end{align}
where $\varphi_t \in \Aut{G}$ is the automorphism appearing in \eqref{eq:Flow_of_GroupAffine_VectorField}. ($T_t \in \Aut{\frk{g}}$ means that $T_t$ is an invertible linear map on $\frk{g}$ that preserves brackets.) Using the automorphism property of $\varphi_t$,
\begin{align*}
\psi_t (g_0) = \psi_t (e^{X_0} \mu_0) = \varphi_t (e^{X_0} \mu_0) \nu_t = \varphi_t (e^{X_0}) \varphi_t (\mu_0) \nu_t = e^{T_t (X_0)} \psi_t (\mu_0) .
\end{align*}
By definition, $\psi_t$ satisfies $\frac{\dd}{\dd t} \psi_t (g_0) = f(\psi_t (g_0))$ and $\lim_{t \rightarrow 0} \psi_t (g_0) = g_0$. Therefore, by time-differentiating the identity $\psi_t (g_0) = e^{T_t (X_0)} \psi_t (\mu_0)$ and using the definition of the flow, one finds that $T_t \in \Aut{\frk{g}}$ must satisfy the ODE
\begin{align}\label{eq:22}
J_{\frk{g}} (T_t (X_0)) \dot{T}_t (X_0) = e^{-T_t (X_0)} f (e^{T_t (X_0)}) - f (I) ,
\end{align}
which is precisely \eqref{eq:ODE_xi_groupaffine}. Since $T_t \in \Aut{\frk{g}}$, there exists a derivation $D : \frk{g} \rightarrow \frk{g}$ such that $T_t (X) = e^{t D} X$. This observation together with \eqref{eq:22} implies that $f$ satisfies
\begin{align}
f(e^X) = e^X J_{\frk{g}}(X) D X + e^X f(I) . 
\end{align}

For the reverse direction, suppose $f$ has the form \eqref{eq:GroupAffineCharacterization} for some derivation $D : \frk{g} \rightarrow \frk{g}$. It is readily verified that \eqref{eq:GroupAffineCharacterization} satisfies \eqref{eq:GroupAffineCondition} when $D = 0$. Therefore, without loss of generality, take $Y_1 = Y_2 = 0$. Let $X_1 , X_2 \in \frk{g}$, and let $\BCH{X_1}{X_2} \in \frk{g}$ be the element of $\frk{g}$ such that
\begin{align}
e^{\BCH{X_1}{X_2}} = e^{X_1} e^{X_2} . 
\end{align}
In other words, $\BCH{X_1}{X_2} \equiv \log{e^{X_1} e^{X_2}}$. As is well-known and suggested by our notation, $\BCH{X_1}{X_2}\in\frk{g}$ is the BCH series, which consists of a series of nested commutators of $X_1$ and $X_2$.  With this context, the left hand side (LHS) of \eqref{eq:GroupAffineCondition} reads
\begin{align}\label{eq:Proof_of_GA_LeftSide}
f(e^{X_1} e^{X_2}) = e^{X_1} e^{X_2} J_{\frk{g}}(\BCH{X_1}{X_2}) D \hspace{0.5mm}\BCH{X_1}{X_2} ,
\end{align}
while the RHS of \eqref{eq:GroupAffineCondition} reads
\begin{align}\label{eq:Proof_of_GA_RightSide}
e^{X_1} f(e^{X_2})  + f(e^{X_1}) e^{X_2} - e^{X_1} f(I) e^{X_2} = e^{X_1} e^{X_2} J_{\frk{g}}(X_2) D X_2 + e^{X_1} J_{\frk{g}}(X_1) D X_1 e^{X_2}  .
\end{align}
Hence, the objective is to argue that the following equality holds:
\begin{align*}
e^{X_1} e^{X_2} J_{\frk{g}}(\BCH{X_1}{X_2}) D \hspace{0.5mm} \BCH{X_1}{X_2} = e^{X_1} e^{X_2} J_{\frk{g}}(X_2) D X_2 + e^{X_1} J_{\frk{g}}(X_1) D X_1 e^{X_2} .
\end{align*}

Notice that $e^{\BCH{X_1}{X_2}} J_{\frk{g}}(\BCH{X_1}{X_2}) D \hspace{0.5mm} \BCH{X_1}{X_2} = \frac{\dd}{\dd t} e^{e^{t D} \BCH{X_1}{X_2}} \Big|_{t = 0}$. Since $D$ is a derivation, $e^{t D} \in \Aut{\frk{g}}$. Using the fact that $\BCH{X_1}{X_2}$ is given by a series of nested commutators, $e^{t D} \BCH{X_1}{X_2} = \BCH{ e^{t D}  X_1}{  e^{t D}  X_2} $, and so
\begin{align*}
e^{\BCH{X_1}{X_2}} J_{\frk{g}}(\BCH{X_1}{X_2}) D \hspace{0.5mm} \BCH{X_1}{X_2} & = \frac{\dd}{\dd t} e^{\BCH{ e^{t D}  X_1}{  e^{t D}  X_2}} \Big|_{t = 0} \\
& = \frac{\dd}{\dd t} e^{e^{tD} X_1} e^{ e^{tD} X_2} \Big|_{t = 0} \\
& = e^{X_1} J_{\frk{g}}(X_1) D X_1 e^{X_2} + e^{X_1} e^{X_2} J_{\frk{g}}(X_2) D X_2 . 
\end{align*}
The identity follows. $\square$

\section{The class of CG distributions}\label{sec:CG}

The definition of a CG on a (matrix) Lie group has been alluded to at several points in the preceding exposition. It is now appropriate to present a formal definition. First, consider a "matrization" map $\cal{M}_{\frk{g}} : \R^{\dim{\frk{g}}} \rightarrow \frk{g}$ that associates an element of $\R^{\dim{\frk{g}}}$ to its matrix representation in $\frk{g}$ with respect to some basis for $\frk{g}$ (c.f.~eqn.~\eqref{def:Ms}). Let $\xi \sim \cal{N}^n \left(0, \Sigma \right)$ be an $n$-dimensional Gaussian distributed vector with zero mean and covariance matrix $\Sigma$. Fix a group element $\mu \in G$, which is to be interpreted as the MAP estimate (this may also be interpreted as the \textit{Lie group mean} \cite{wolfe2011bayesian}). From $\xi$ and $\mu$, induce a random element $g \in G$ via the formula
\begin{align}\label{def:CG}
g = e^{\M{\frk{g}}{\xi}} \mu .
\end{align}
Write $g \sim \cal{T}_G ( \mu , \Sigma )$ to indicate $g$ is a random element of $G$ with a distribution following \eqref{def:CG}. The distribution $\cal{T}_G ( \mu , \Sigma )$ on $G$ induced via eqn.~\eqref{def:CG} is known as a (left) CG in the literature \cite{wolfe2011bayesian,barrau2014intrinsic,brossard2017unscented}. A \textit{right} CG has the $e^{\M{\frk{g}}{\xi}}$ and $\mu$ in \eqref{def:CG} interchanged. A CG has the physically intuitive interpretation that there is a reference group element $\mu \in G$ representing the "best" estimate of the state and the uncertainty in the estimate is given by random ``rotations" $e^{\M{\frk{g}}{\xi}} \in G$ generated by the random vector $\xi \in \R^n$. When the Lie group in question is $(\R^n , +)$, then a CG is simply a Gaussian and the theory reduces to the standard Gaussian-based filtering framework on $\R^n$.

There is a decent volume of literature focused on proposing distributions on Lie groups for the purposes of filtering, and no attempt is made to provide a comprehensive review here (see, e.g., \cite{AttitudeSurvey07,barrau2014intrinsic,WangLee2021} for relatively recent reviews and approaches for $\S^3$ and $\SO{3}$). The class of CGs has been considered by other authors for the purposes of estimation on general Lie groups and, in particular, attitude estimation \cite{barfoot2014associating,barrau2014intrinsic,brossard2017unscented}. However, one major contribution of this manuscript, in addition to Theorem \ref{thm:Characterization}, is the following corollary of Theorem \ref{thm:Characterization}.
\begin{cor}\label{cor:1}
Let $g_0 \sim \cal{T}_G ( \mu_0 , \Sigma_0 )$ for some $\mu_0 \in G$ and positive-definite, symmetric $\Sigma_0 \in \R^{n \times n}$. Consider the following IVP on $G$
\begin{align*}
\left\lbrace \begin{array}{l}
\dot{g} = f (g) \\
g(0) = g_0 \in G .
\end{array} \right. 
\end{align*}
Suppose $f$ is group-affine (i.e., it satisfies \eqref{eq:GroupAffineCondition}). Then, for all $t > 0$, the solution $g : [0 , \infty) \rightarrow G$ of the IVP satisfies $g (t) \sim \cal{T}_G ( \mu (t) , \Sigma (t) )$, where $\mu : [0 ,\infty) \rightarrow G$ is the solution of the IVP
\begin{align*}
\left\lbrace \begin{array}{l}
\dot{\mu} = f (\mu) \\
\mu(0) = \mu_0 \in G ,
\end{array} \right. 
\end{align*}
and $\Sigma : [0 , \infty) \rightarrow \R^{n \times n}$ solves
\begin{align*}
\left\lbrace \begin{array}{l}
\dot{\Sigma} = ( D + \ad_{Y_2} ) \Sigma + \Sigma ( D + \ad_{Y_2} )^T \\
\Sigma (0) = \Sigma_0 \in \R^{n \times n} ,
\end{array} \right. 
\end{align*}
where $D$ and $Y_2$ are defined via \eqref{eq:GroupAffineCharacterization}. 
\end{cor}

Corollary \ref{cor:1} demonstrates that the class of CGs \eqref{def:CG} is invariant under the flow generated by group-affine dynamics. The somewhat surprising aspect of this is that group-affine dynamics may be nonconstant (on the group), yet invariance is maintained (see, e.g., Section \ref{sub:SE23}). For conventional Kalman filtering, it is well-known that linear state space dynamics take Gaussian initial conditions to Gaussians with updated parameters. What has just been proven is that for CGs the analogous condition on the vector field is that it is group-affine. This observation provides a case for the claim that CGs are the natural class of probability distributions on a Lie group to filter with respect to, assuming the dynamics under consideration are group-affine. (It is being assumed that the posterior distribution after an observation on the state is well approximated by a CG when making this claim. This may not always be the case.) In fact, given an underlying state space (i.e., a manifold of some dimension $ \geq 1$) and a dynamical system on this state space, one is encouraged to design a group law which makes the associated vector field group-affine (c.f.~\cite{barrau2022geometry}). If this is not possible, and one still insists on filtering with respect to CGs, then one should seek a group law which makes the dynamics "as close to group-affine" as possible. Numerical evidence for this claim is provided in the special case of attitude filtering in the presence of gyro bias (see Section \ref{sec:GyroBias}). 

\section{The TSSDE}\label{sec:TSSDE}

In this section, a stochastic generalization of a group-affine dynamical system as required for filtering is discussed. For simplicity, the focus is on the case where the deterministic part of the dynamics is constant (i.e., assuming $D = 0$ in \eqref{eq:GroupAffineCharacterization}). The more general case carries over straightforwardly with minimal modifications to the calculations below.
 
Suppose there is a smooth function $g : [0 ,\infty) \rightarrow G$ that satisfies
\begin{align}\label{eq:PoissonLieGroup}
\left\lbrace \begin{array}{l}
\dot{g} (t) = X (t) g (t) \\[3pt]
g(0) = g_0 \in G .
\end{array} \right. 
\end{align}
for some (possibly time-dependent) $X (t) \in \frk{g}$. The solution to \eqref{eq:PoissonLieGroup} may be written in a Lie-algebraic way as $g(t) = e^{\Theta (t)} g_0$ with $\Theta (t) \in \frk{g}$ satisfying $\left( \int_0^1 e^{s \ad_{\Theta (t)}} \dd s \right) \Theta' (t) = X (t) $ and giving rise to the well-known Magnus expansion. In particular, if $[X(t_1) , X(t_2)] = 0$ for all $t_1 , t_2 \geq 0$, then $g(t) = e^{X(t)} g_0$.

Let $\eta : \R \rightarrow \R^n$ be a Brownian motion with delta-correlated spectral density matrix $Q_{\eta}$. Consider the following Langevin equation on $G$:
\begin{align}\label{eq:LangevinEquationG}
\left\lbrace \begin{array}{l}
\dot{g} = ( X + \M{\frk{g}}{\eta} ) g \\[3pt]
g(0) = g_0 \in G .
\end{array} \right. 
\end{align}
As an SDE, \eqref{eq:LangevinEquationG} must carry the Stratonovich interpretation, otherwise solutions may not remain in $G$. Suppose $g$ satisfying \eqref{eq:LangevinEquationG} is of the form $g (t) = e^{\M{\frk{g}}{\xi (t)}} \mu (t)$, where $\mu (t) \in G$ is given by $\mu (t) = e^{\Theta (t)} g_0$ and the equation for $\xi (t) \in \R^n$ is to-be-determined. In what follows, drop the implied $t$-dependence. Following a derivation similar to that of eqn.~\eqref{eq:ODE_xi}, $\dot{\xi}$ satisfies
\begin{align}\label{eq:Derivation_ConstantDynamics_1}
J_{\frk{g}}(\M{\frk{g}}{\xi}) \M{\frk{g}}{\dot{\xi}} = e^{-\M{\frk{g}}{\xi}} ( X + \M{\frk{g}}{\eta} ) e^{\M{\frk{g}}{\xi}} - X . 
\end{align}
Applying the braiding identity \eqref{eq:braiding} to \eqref{eq:Derivation_ConstantDynamics_1} and simplifying yields
\begin{align}\label{eq:Derivation_ConstantDynamics_2}
J_{\frk{g}}(\M{\frk{g}}{\xi}) \M{\frk{g}}{\dot{\xi}} = \left( e^{- \ad_{\M{\frk{g}}{\xi}}} - I \right) X + e^{- \ad_{\M{\frk{g}}{\xi}}} \M{\frk{g}}{\eta} .
\end{align}
To simplify eqn.~\eqref{eq:Derivation_ConstantDynamics_2} further, introduce the operator $W_{\!\frk{g}} : \frk{g} \rightarrow \R^{n\times n}$ given by
\begin{align} \label{def:Wg}
W_{\!\frk{g}} (X) = \left( \int_0^1 e^{s \ad_X} \dd s \right)^{-1} ,
\end{align}
and note $J_{\frk{g}}^{-1} (X) e^{- \ad_X} = W_{\!\frk{g}} (X)$. Noting further that $J_{\frk{g}}^{-1} (X) (e^{- \ad_X} - I) = - \ad_X$, eqn.~\eqref{eq:Derivation_ConstantDynamics_2} simplifies to
\begin{align}\label{eq:Derivation_ConstantDynamics_3}
\M{\frk{g}}{\dot{\xi}} = \ad_X \M{\frk{g}}{\xi} + W_{\!\frk{g}}(\M{\frk{g}}{\xi}) \M{\frk{g}}{\eta} .
\end{align}

Eqn.~\eqref{eq:Derivation_ConstantDynamics_3} is essentially our desired SDE defined on the Lie algebra. However, it is an equation for matrices and it is desirable to convert this to an equation for the coordinates $\xi \in \R^n$. In order to accomplish this, introduce the map $\cal{V}_{\frk{g}} : \frk{g} \rightarrow \R^n$ defined as the inverse of $\cal{M}_{\frk{g}} : \R^n \rightarrow \frk{g}$. Using this map, define 
\begin{align}\label{def:adbar}
\adbar_{\xi_1} \xi_2 := \V{\frk{g}}{ \ad_{\M{\frk{g}}{\xi_1}} \M{\frk{g}}{\xi_2} } ,
\end{align}
for $\xi_1 , \xi_2 \in \R^n$. For each $\xi \in \R^n$, $\adbar_{\xi} : \R^n \rightarrow \R^n$ is a linear map and, hence, may be represented as a matrix. Examples of this matrix representation of the adjoint are in Section \ref{sec:Examples}. Moreover, the identity $\M{\frk{g}}{\adbar_{\xi_1} \xi_2} = \ad_{\M{\frk{g}}{\xi_1}} \M{\frk{g}}{\xi_2}$ holds for all $\xi_1 , \xi_2 \in \R^n$. This identity is a consequence of the fact that the maps $\cal{V}_{\frk{g}}$ and $\cal{M}_{\frk{g}}$ are inverses of each other. From this identity, introduce
\begin{align}\label{def:Wgbar}
\overline{W}_{\!\!\frk{g}} (\xi) := \left( \int_0^1 e^{s \adbar_{\xi}} \dd s \right)^{-1} 
\end{align}
and note that $W_{\!\frk{g}}(\M{\frk{g}}{\xi}) \M{\frk{g}}{\eta} = \M{\frk{g}}{\overline{W}_{\!\!\frk{g}} (\xi) \eta}$. Using these notations, one may "vectorize" \eqref{eq:Derivation_ConstantDynamics_3} to find the equation
\begin{align}\label{eq:TSSDE}
\dot{\xi} = \adbar_{\V{\frk{g}}{X}} \xi + \overline{W}_{\!\!\frk{g}} (\xi) \eta .
\end{align}
Again, as an SDE, eqn.~\eqref{eq:TSSDE} must carry the Stratonovich interpretation. If the derivation $D$ in eqn.~\ref{eq:GroupAffineCharacterization} is non-zero, then \eqref{eq:TSSDE} becomes
\begin{align}\label{eq:TSSDE_2}
\dot{\xi} = (\overline{D} + \adbar_{\V{\frk{g}}{X}}) \xi + \overline{W}_{\!\!\frk{g}} (\xi) \eta ,
\end{align}
where $\overline{D} : \R^n \rightarrow \R^n$ is the linear map defined by $\overline{D} (\xi) := \V{\frk{g}}{D \M{\frk{g}}{\xi}}$ for $\xi \in \R^n$.

Eqn.~\eqref{eq:TSSDE_2} represents the TSSDE relevant for propagation of the group elements in terms of the \emph{exponential coordinates} which parameterize the Lie algebra through the vectorization operation. Clearly, if $\eta = 0$, then \eqref{eq:TSSDE_2} becomes a linear equation, which is consistent with Theorem \ref{thm:Characterization}. Due to the nonlinear, state-dependent diffusion tensor $\overline{W}_{\!\!\frk{g}} (\xi)$ in \eqref{eq:TSSDE_2}, it no longer appears as though the class of CGs are invariant under propagation in the presence of noise. Indeed, at the end of Section \ref{sub:SU2}, the FPE associated with \eqref{eq:TSSDE_2} for the case of $\S^3$ (i.e., the attitude filtering problem in the absence of gyro bias) is analyzed and the fundamental solution of this FPE is explicitly computed. The FPE fundamental solution defines the class of probability distributions which remain invariant during propagation in the presence of noise, and the computed fundamental solution for $\S^3$ is notably \textit{not} a CG. It is shown, however, that this fundamental solution is well-approximated by a CG when $\Delta t Q_{\eta} \ll 1$ (in appropriate units). In many applications $\Delta t Q_{\eta}$ is small and the CG approximation appears to hold with high accuracy during propagation. 

\section{Filtering on Lie groups: the TSF}\label{sec:Filtering_on_Lie_groups}

With the TSSDE in hand, one has at their disposal the myriad of nonlinear filtering techniques to determine how the mean and covariance of $\xi$ (approximately) propagate through \eqref{eq:TSSDE_2}. The perhaps crudest approach would involve linearization in an EKF fashion and then proceed to numerically integrate the resulting ODEs defining the (approximate) mean and covariance evolution. One can do much better, however. In Section \ref{sub:propagation} a generalization of the CTUT described in \cite{sarkka2007unscented} to the case of state-dependent diffusion is provided. It is used to propagate the mean and covariance of $\xi$ through \eqref{eq:TSSDE_2}. This approach provides a relatively accurate way to update the mean and covariance of $\xi$ in between measurement updates within the TSF framework.

Propagation of distributional parameters is only one of two fundamental stages in recursive filtering, with the other being the measurement stage, i.e. refining these distributional parameters by an incoming measurement $z$ according to Bayes' theorem. In this regard, a measurement $z \in \R^d$ is assumed to be related to the group element $g$ through the equation $z = h(g,w)$, where $h : G \times \R^m \rightarrow \R^d$ is a function of the group element and a random vector $w \in \R^m$ that is typically assumed to be $w \sim \cal{N}^m (0,R)$. The function $h$ associating group elements to measurement values is typically of a form that demands an approximate means of performing the Bayesian fusion. In this work, we advocate for utilization of the UT \cite{UnscentedKF04}. Its application to the measurement stage of the TSF is described Section \ref{sub:MeasurementUpdate}. A similar approach to a UT-based measurement update on Lie groups using a CG as a reference distribution appeared in \cite{brossard2017unscented,sjoberg2021lie}.

When using the CTUT to update the first two moments of the distribution over $\xi$ according to the TSSDE \eqref{eq:TSSDE_2}, the mean of $\xi$ will no longer remain zero, even if it is initially zero. This is a result of the nonlinear, state-dependent diffusion tensor $\overline{W}_{\!\!\frk{g}} (\xi)$ present in \eqref{eq:TSSDE_2}. (However, it has been observed that the mean generally remains small over time scales $\Delta t$ for which $\Delta t \sigma \ll 1$, where $\sigma$ is the largest eigenvalue of $Q_{\eta}$). The mean of $\xi$ will also generally become non-zero after the measurement stage. Because the definition of a CG \eqref{def:CG} requires the mean of $\xi$ to be zero (in order to interpret $\mu$ as the MAP estimate), a procedure to approximate the distribution $g = e^{\M{\frk{g}}{\xi}} \mu$, $\xi \sim \cal{N}^n ( \wh{\xi} , \Sigma )$, by a CG must be developed. This is done in Section \ref{sub:Whitening} and the resulting procedure is called \textit{whitening}. Whitening is fundamental to the TSF and is required to produce statistically consistent MAP estimates.  

Since the bulk of the TSF algorithm presented here relies on the UT, it is helpful to briefly review its basics. The UT represents the distribution of a random vector $x \in \R^n$ as a collection of \textit{sigma points} $\sigma_i \in \R^n$, $i = 0 , \cdots , 2 n$, given by
\begin{align}\label{eq:sigmapts}
\left\lbrace \begin{array}{ll}
\sigma_0 = \wh{x} & \\[5pt]
\sigma_i = \wh{x} + \sqrt{n+\lambda} ~ L_{: i} & i = 1 , \cdots n \\[5pt]
\sigma_i = \wh{x} - \sqrt{n+\lambda} ~ L_{: i} & i = n+1 , \cdots, 2n ,
\end{array} \right.
\end{align}
in which $L_{:i}$ denotes the $i^{\rm{th}}$ column of the Cholesky decomposition of the covariance matrix of $x$, and $\lambda \neq - n$ is a tuning parameter that controls the spread of the sigma points around their mean. In addition, one defines the collection of weights
\begin{align}\label{eq:UT_weight}
\left\lbrace \begin{array}{lc}
w_0 = \dfrac{\lambda}{\lambda+n} & \\[5pt]
w_i = \dfrac{1}{2\left(\lambda+n\right)} , & i = 1, \cdots, 2n .
\end{array} \right.
\end{align}
The unscented transform then approximates the expectation of the distribution of $y = f(x)$ for an arbitrary map $f : \R^n \rightarrow \R^m$ as $\E{ f(x) } \simeq \sum_{i=0}^{2n} w_i f\left(\sigma_i\right)$. There are variants of the UT, such as the simplex UT, that use fewer sigma points to represent the distribution of $x$, thereby reducing the computational resources needed to evaluate summations. In this manuscript, only the "vanilla" UT as just described is considered.

\subsection{Propagation}\label{sub:propagation}

This section is devoted to describing the propagation step within the TSF using the CTUT \cite{sarkka2007unscented}. However, the CTUT within \cite{sarkka2007unscented} did not consider state-dependent diffusion, as the $\overline{W}_{\!\!\frk{g}}(\xi) \dd \eta$ term in \eqref{eq:TSSDE_2} now demands. Hence, it is necessary to derive a generalization to \cite[Algorithm 4.4]{sarkka2007unscented}. This is done for a general (Stratonovich) SDE
\begin{align}\label{eq:SDE_stacked} 
\dd x = f(x) \dd t+ G(x) \dd w ,
\end{align}
where $x \in \R^n$ is a vector-valued stochastic process, $f : \R^n \rightarrow \R^n$ is a vector field governing the deterministic portion of the dynamics of $x$, $G : \R^n \rightarrow \R^{n \times m}$ is the diffusion tensor, and $w$ is a Brownian motion with delta-correlated spectral density $Q \in \R^{m \times m}$. Eqn.~\eqref{eq:TSSDE_2} is reproduced by setting $\xi \to x$, $\eta \to w$, $f(x) = \overline{D} + \adbar_{\V{\frk{g}}{X}} x$, and $G(x) = \overline{W}_{\!\!\frk{g}} (x)$. Interpreting \eqref{eq:SDE_stacked} as a Stratonovich SDE, the FPE for the time-dependent transition probability density $p : \R \times \R^n \rightarrow [ 0 , \infty )$ associated with \eqref{eq:SDE_stacked} reads
\begin{align}\label{eq:FPE_general}
\partial_t p = - \diver{ f p } + \frac{1}{2} \sum_{i = 1}^n \sum_{k , \ell = 1}^m \sum_{j = 1}^n \partial_i ( G_{ik} Q_{k \ell} \partial_j ( G_{j \ell} p ) ) .
\end{align}
It is possible to rewrite this equation as
\begin{align}\label{eq:FPE_general_Ito}
\partial_t p = - \diver{ \tilde{f} p } + \frac{1}{2} \sum_{i,j = 1}^n \partial_i \partial_j ( ( G Q G^T )_{ij} p ) ,
\end{align}
where $\tilde{f} : \R^n \rightarrow \R^n$ is a vector field whose $i$th component is
\begin{align}\label{eq:tildef_def}
\tilde{f}_i (x) = f_i (x) + \frac{1}{2} \sum_{k, \ell = 1}^m \sum_{j = 1}^n G_{j \ell} Q_{k \ell} \partial_j G_{ik} . 
\end{align}
By multiplying \eqref{eq:FPE_general_Ito} by $x$ and $x x^T$ and integrating, one may derive the \textit{moment evolution equations} for the mean $\wh{x} := \E{x}$ and covariance $P := \Cov{x} = \E{(x-\wh{x})(x-\wh{x})^T}$ corresponding to \eqref{eq:FPE_general_Ito}:
\begin{align}\label{eq:MomentEvolution_stacked}
\left\lbrace \begin{array}{l}
\dfrac{\dd \wh{x}}{\dd t} = \E{ \tilde{f} (x) } \\[5pt] 
\dfrac{\dd P}{\dd t} = \Cov{ x , \tilde{f}(x) } + \Cov{ \tilde{f} (x) , x } + \E{ G (x) Q G^T (x) } ,
\end{array} \right.
\end{align}
where $\Cov{ x,f(x) } = \E{(x - \wh{x})(f(x) - \overline{f})^T}$.

As discussed in \cite[Appendix]{sarkka2007unscented}, the CTUT propagation equations in \cite[Algorithm 4.4]{sarkka2007unscented} are obtained by applying the UT to the expectations on the RHS of equations \eqref{eq:MomentEvolution_stacked}. However, as mentioned earlier, that derivation did not consider state-dependent $G$ in \eqref{eq:MomentEvolution_stacked}. Hence, the covariance evolution in \cite[Algorithm 4.4]{sarkka2007unscented} must be augmented  with the term $\E{ G (x) Q G^T (x) } \simeq \sum_{i = 0}^{2n} w_i G ( \sigma_i ) Q G^T ( \sigma_i )$, where the weights $w_i$ are defined in \eqref{eq:UT_weight}. With this, all the needed ingredients are present to write down the system of coupled ODEs governing the evolution of the sigma points during the propagation step of the TSF. The derivation follows that in \cite[Appendix]{sarkka2007unscented} closely and is omitted. The end result is essentially \cite[Equation 35]{sarkka2007unscented} with the matrix $M$ therein modified to include the term $\sum_{i = 0}^{2n} w_i G ( \sigma_i ) Q G^T ( \sigma_i )$.

Depending on the dimension of $G$ and the technique used to numerically integrate the sigma point ODEs, it may be too computationally expensive to use the CTUT for real-time applications. In this regard, another approach for moment propagation through the TSSDE is briefly indicated. This approach is based on a "short-time expansion" of the FPE and may be seen as a generalization of the EKF to state-dependent diffusion. Assuming the initial probability density associated with $\xi$ in \eqref{eq:TSSDE_2} is sufficiently localized around the origin of the Lie algebra, then a physically reasonable approximation is to assume it remains localized around the origin over propagation time scales $\Delta t$ for which $q_{\!\eta} \Delta t \ll 1$ (in appropriate units), where $q_{\!\eta}$ is the largest eigenvalue of $Q_{\eta}$, the spectral density associated with $\eta$ in \eqref{eq:TSSDE_2}. Thus, only the behavior of the diffusion tensor $\overline{W}_{\!\!\frk{g}}$ near the origin contributes to the evolution of the density under the FPE. More rigorously, for each $\lambda > 0$, consider a rescaled density $p_{\lambda} (t , x) = \lambda^{-n} p (t,x/\lambda)$, where $p(t,x)$ is assumed to satisfy \eqref{eq:FPE_general_Ito}. Rewriting the FPE \eqref{eq:FPE_general_Ito} for $p_{\lambda}$, one may expand the coefficients of this transformed PDE in powers of $\lambda$. Removing terms that fall off faster than $\lambda^{-1}$, one arrives at an "effective" FPE that captures the evolution of the density over "short" time scales. The result of this procedure is stated for the $\SU{2}$ example in Section \ref{sub:SU2}. Similar ideas are described in \cite{ye2023uncertainty,ye2024uncertainty}.

\subsection{Measurement update}\label{sub:MeasurementUpdate}

The previous section was concerned with the prediction stage of the TSF. The present section is concerned with the measurement stage of filtering, in which our state estimates and their uncertainties are refined by an incoming measurement $z$ according to Bayes' theorem. The measurement $z \in \R^d$ is assumed to be related to the group element $g$ through the equation $z = h(g) + w$, where $h : G \rightarrow \R^d$ and $w \in \R^d$ is assumed to be a zero-mean Gaussian distributed random vector. An example measurement model in the case of attitude filtering is a triaxial magnetometer which measures an ambient magnetic field $B$ in the coordinate frame to which the attitude is referenced. Rotating to the body frame coordinates using eqn.~\eqref{eq:Rorth}, $h(q) = R(q) B$, where $R(q)$ is given by \eqref{eq:Rorth}. Note that this requires a model of the ambient magnetic field $B$.

Rarely will $h$ be equivalent to a linear function on the Lie algebra of $G$. This means the class of CGs over the group are not, in general, invariant under the Bayesian update. Hence, nonlinear filtering techniques are required to approximate the posterior distribution over the group. To apply the UT to the measurement stage of filtering using a CG \eqref{def:CG} as the reference distribution, start by generating sigma points according to \eqref{eq:sigmapts}, where the random vector $\xi \in \R^n$ in \eqref{def:CG} is used in place of $x$ in \eqref{eq:sigmapts}. These sigma points are then transformed to "measurement space" via $\nu^i = h \left( e^{\M{\frk{g}}{\sigma^i}} \mu \right)$. Since both the measurement space and the Lie algebra where the sigma points are defined have the structure of a finite-dimensional vector space, the Bayesian fusion amounts to a standard Kalman filter update. These update equations are standard and omitted for brevity (see, e.g., \cite[Eqns.~2-4]{Crassidis_UKF}). This procedure will produce an (approximate) posterior distribution over $G$ of the form $g = e^{\M{\frk{g}}{\xi}} \mu$, where $\xi \sim \cal{N}^n ( \wh{\xi} , \Sigma )$ and $\wh{\xi} \neq 0$ in general. In order to maintain consistency, this posterior must be transformed into the form of a CG. This process is called whitening and is described in the following Section. 

\subsection{Whitening}\label{sub:Whitening}

In this section, a procedure is developed to approximate the distribution $g = e^{\M{\frk{g}}{\xi}} \mu$, with $\xi \sim \cal{N}^n ( \wh{\xi} , \Sigma )$, by a CG \eqref{def:CG}. In contrast to eqn.~\eqref{def:CG}, $\wh{\xi}$ is not assumed to be zero. The approximation is accomplished by rotating $\mu$ by a predetermined amount, while applying the reverse rotation to the distribution $\xi$, where the rotation is chosen in such a way that the new distribution over $\xi$ has zero mean to some specified tolerance. The resulting algorithm will be referred to as \textit{whitening}. 

The problem is to find a fixed vector $a \in \R^n$ such that the mean of
\begin{align}\label{eq:tilde_xi}
\tilde{\xi} := \V{\frk{g}}{ \BCH{ \M{\frk{g}}{\xi} }{ -\M{\frk{g}}{a} } }
\end{align}
is zero. If this can be accomplished, then write
\begin{align*}
g = e^{\M{\frk{g}}{\xi}} \mu = \left( e^{\M{\frk{g}}{\xi}} e^{-\M{\frk{g}}{a}} \right) e^{\M{\frk{g}}{a}} \mu = e^{\M{\frk{g}}{\tilde{\xi}} } \tilde{\mu} ,
\end{align*}
where $\tilde{\mu} = e^{\M{\frk{g}}{a}} \mu$. Since $\tilde{\xi}$ has zero mean, $\tilde{\mu}$ may be interpreted as the MAP of the distribution over $G$ in accordance with the definition of a CG \eqref{def:CG}. Due to the nonlinear nature of the BCH expansion, it seems improbable that one may find a closed-form for an $a$ in terms of the moments of $\xi$. Instead, a simpler route is taken that can be iterated upon until a desired convergence criteria is met. In this vein, an $a \in \R^n$ is sought such than the mean of $\tilde{\xi}$ is strictly \textit{smaller} (in magnitude) that the mean of $\xi$. 

To proceed, first recall a few facts about the BCH expansion. Let $X,Y \in \frk{g}$, where $\frk{g}$ is the Lie algebra of some (matrix) Lie group $G$. Consider the series representation of BCH \cite[Section 5.6]{hall2013lie}, whose first few terms are given by
\begin{align}\label{eq:BCH_series}
\BCH{X}{Y} = X + Y + \frac{1}{2} [X,Y] + \frac{1}{12} ([X,[X,Y]] - [Y,[X,Y]]) + \cdots . 
\end{align}
If all the commutators in \eqref{eq:BCH_series} that only contain a single $Y$ are collected, then it is possible to write an alternative form of the BCH series given by
\begin{align}\label{eq:BCH_series_2}
\BCH{X}{Y} = X + J_{\frk{g}}^{-1}(X) Y + \cal{O} (Y^2) ,
\end{align}
where $J_{\frk{g}}(X)$ was introduced in \eqref{eq:JacobianExpMap}.

Using formula \eqref{eq:BCH_series_2} to expand the RHS of \eqref{eq:tilde_xi} and vectorizing, one finds
\begin{align}\label{eq:tilde_xi_expansion}
\tilde{\xi} = \xi - \overline{J}^{-1}_{\frk{g}} (\xi) a + \cal{O} (a^2) ,
\end{align}
where $\overline{J}_{\!\frk{g}}$ is defined similar to \eqref{def:Wgbar}. If one takes the expectation of both sides of \eqref{eq:tilde_xi_expansion}, and imposes the constraint that $\E{\tilde{\xi}} = 0$, then
\begin{align}\label{eq:tilde_xi_expansion_mean}
0 = \wh{\xi} -  \E{ \overline{J}_{\!\frk{g}}^{-1}(\xi) } a + \cal{O} (a^2) .
\end{align}
Ignoring higher-order terms in $a$, solve \eqref{eq:tilde_xi_expansion_mean} for $a$ to find
\begin{align}\label{eq:approximate_soln_a}
a \simeq \left( \E{ \overline{J}_{\!\frk{g}}^{-1}(\xi) } \right)^{-1} \wh{\xi} .
\end{align}

In order to use \eqref{eq:approximate_soln_a}, one must be able to compute $\E{ \overline{J}_{\!\frk{g}}^{-1}(\xi) }$ when $\xi \sim \cal{N}^n ( \wh{\xi} , \Sigma )$. Without making simplifying assumptions about the covariance $\Sigma$ of $\xi$, it appears difficult to find a closed form. There are at least two routes to approximating this expectation value. One route is to use the UT, while another may use a first-order Taylor series approach, i.e. $\E{ \overline{J}_{\!\frk{g}}^{-1}(\xi) } \simeq \overline{J}_{\!\frk{g}}^{-1}(\wh{\xi})$. Using the fact that $\overline{J}_{\!\frk{g}}^{-1}(\xi) \xi = \xi$ for any $\xi \in \R^n$, then under this first-order approximation, $a \simeq \wh{\xi}$. However, if $a$ in \eqref{eq:approximate_soln_a} is computed using, for example, the UT, then, in general, $a \neq \wh{\xi}$.

To compute the covariance of $\tilde{\xi}$, expand \eqref{eq:tilde_xi} as a series in $\xi$, not $a$. For this, note the BCH symmetry $\BCH{X}{Y} = - \BCH{-Y}{-X}$, true for any $X,Y \in \frk{g}$. Then, there is an expansion similar to \eqref{eq:tilde_xi_expansion}, but now in powers of $\xi$:
\begin{align}\label{eq:tilde_xi_expansion_alternate}
\tilde{\xi} = \overline{J}_{\!\frk{g}}^{-1}\left(a\right) \left(\xi - a\right) + \cal{O} (\xi^2) .
\end{align}
The previous expression tells us that the covariance of $\tilde{\xi}$ is $\tilde{\Sigma} \simeq \overline{J}_{\!\frk{g}}^{-1} \left(a\right) ( \Sigma + (a - \wh{\xi})(a - \wh{\xi})^T ) \overline{J}_{\!\frk{g}}^{-T} \left(a\right)$. Choose $a = \wh{\xi}$. Then $\tilde{\Sigma} \simeq \overline{J}_{\!\frk{g}}^{-1} \left(\wh{\xi}\right) \Sigma \overline{J}_{\!\frk{g}}^{-T} \left(\wh{\xi}\right)$. 

If $a$ is computed based on an approximation such as \eqref{eq:approximate_soln_a}, then the mean of $\tilde{\xi}$ in \eqref{eq:tilde_xi} will not be \textit{exactly} zero. However, by analyzing higher-order terms in \eqref{eq:tilde_xi_expansion_alternate} using, say, $a = \wh{\xi}$, one may show that the new mean of $\tilde{\xi}$ has norm which is strictly \textit{smaller} in norm than $\wh{\xi}$. This implies that if the mean and covariance of $\tilde{\xi}$ are (approximately) computed using, for example, the choice $a = \wh{\xi}$, then this whitening process may be repeated using this new mean and covariance until a desired convergence criteria is met. In practice, it has been observed that a small number of iterations, e.g. $\mathcal{O}\left(10\right)$, will suffice to produce a distribution with a mean on the order of machine double precision.

\subsection{Summary of the TSF}\label{sub:summary}

Sections \ref{sub:propagation}, \ref{sub:MeasurementUpdate}, and \ref{sub:Whitening} may be summarized to describe a complete TSF based on the UT. One recursive step of the filter proceeds as follows. Let $g \sim \cal{T}_G ( \mu_- , \Sigma_- )$ be the prior CG. In particular, $g = e^{\M{\frk{g}}{\xi_-}} \mu_-$ where $\xi_- \sim \cal{N}^n (0 , \Sigma_-)$. Suppose there is an incoming measurement $z \in \R^d$. The first step is to update the distribution of $\xi_-$ with the measurement $z$ using the UT as described in Section \ref{sub:MeasurementUpdate}. This produces a new distribution $\xi_+ \sim \cal{N}^n ( \wh{\xi}_+ , \Sigma_+ )$ where, in general, $\wh{\xi}_+ \neq 0$. To remedy this, several iterations of the whitening procedure described in Section \ref{sub:Whitening} are performed. This transforms the posterior distribution $g = e^{\M{\frk{g}}{\xi_+}} \mu_-$ into an (roughly) equivalent CG distribution, $g \sim \cal{T}_G ( \tilde{\mu}_+ , \tilde{\Sigma}_+ )$. Next, one must propagate the MAP $\tilde{\mu}_+$ and the first two moments of $\tilde{\xi}_+ \sim \cal{N}^n (0 , \tilde{\Sigma}_+)$ until the next measurement is received. The MAP estimate $\tilde{\mu}_+$ is propagated forward by solving
\begin{align*}
\left\lbrace \begin{array}{l}
\dot{\mu} = f (\mu) \\
\mu(0) = \tilde{\mu}_+ 
\end{array} \right.
\end{align*}
over the necessary time step, where $f$ is the vector field governing the dynamics on the Lie group. The moments of $\tilde{\xi}_+$ are propagated forward in time under the appropriate TSSDE \eqref{eq:TSSDE_2} using the CTUT as described in Section \ref{sub:propagation}. Before the next incoming measurement is processed, another iteration of the whitening procedure is typically required due to the nonlinear, state-dependent diffusion tensor in \eqref{eq:TSSDE_2}. Once this is complete, the algorithm just outlined may be repeated until all measurements are exhausted. 

\subsection{Performance metrics}\label{sub:PerformanceMetrics}

In the context of Gaussian-based nonlinear filtering on $\R^n$, a standard filter performance metric in simulations is the (normalized) $\chi^2$ value. Specifically, if $\chi^2 = \frac{1}{n} \langle x - \mu , P^{-1} (x-\mu) \rangle$ with $x \sim \cal{N}^n (\mu , P)$, then $\chi^2$ follows a chi-squared distribution and has mean $\E{\chi^2} = 1$. If, after averaging over many filter realizations, the $\chi^2$ is substantially different from its theoretical expectation, then there is strong evidence that the distribution followed by the estimated error is \textit{not} a Gaussian with covariance matrix given by the associated state error covariance. Hence, the $\chi^2$ value gives a measure of \textit{statistical consistency} of the filter. Poor statistical consistency could be the result of flawed assumptions, poor parameter tuning, numerical issues, or genuine non-Gaussian behavior (just to name a few). A key aspect of the $\chi^2$ metric is that it takes into account the full state error covariance matrix $P$, capturing important cross-correlations between states. 

The goal here is to identify a similar metric when filtering with respect to CGs on Lie groups. Suppose one is attempting to estimate $g \in G$ and, at any given time in the simulation, an estimate of $g$, denoted $\mu \in G$, is computed and an associated covariance matrix $\Sigma \in \R^{n \times n}$ (here, $n = \dim{G}$). In analogy with filtering on Euclidean space, the error in our estimate $\mu$ is stipulated to be $g \mu^{-1}$. If the TSF's CG approximations of the true underlying probability distribution over $g$ are accurate, then $g \sim \cal{T}_G (\mu , \Sigma)$ and, in particular, $g \mu^{-1} = e^{\M{\frk{g}}{\xi}}$, where $\xi \sim \cal{N}^n (0,\Sigma)$. Hence, $\V{\frk{g}}{ \log{ (g \mu^{-1}) } } \sim \cal{N}^n (0 , \Sigma)$ and so the quantity 
\begin{align}\label{def:chisquareG}
\chi^2_G := \frac{1}{n} \langle \V{\frk{g}}{ \log{ (g \mu^{-1}) } } , \Sigma^{-1} \V{\frk{g}}{ \log{ (g \mu^{-1}) } } \rangle ,
\end{align}
averaged over many random realizations should be (approximately) $1$. The quantity $\chi^2_G$ will be used as the primary metric to evaluate simulated filter performance in Section \ref{sub:experiment}. 

\section{Examples}\label{sec:Examples}

In the following sections, eqn.~\eqref{eq:TSSDE_2} is examined for some specific examples of Lie groups that are relevant for applications. Most of the work lies in determining the explicit form of the diffusion tensor \eqref{def:Wgbar} for the given example. 

\subsection{$\SU{2}$}\label{sub:SU2}

Consider the case of unit quaternions, which define the Lie group $\mathbb{S}^3$ under quaternionic multiplication. This Lie group is isomorphic to $\SU{2}$, hence the name of this example. As discussed in Sec.~\ref{sec:AttEst}, quaternions define a convenient parameterization of elements of $\SO{3}$, which are the primary mathematical objects of interest in attitude filtering. In this case, the relevant group equation of motion is eqn.~\eqref{eq:qEOM}, where $\omega$ is a noisy measurement provided by a gyroscope. This section does not consider the case where $\omega$ is corrupted by a bias, as this is reserved for Section \ref{sec:GyroBias}. Instead, simply assume the measured angular rate satisfies $\omega_m = \omega + \eta$, where $\eta \sim \cal{N}^3 (0 , Q_{\eta})$. Under the paradigm of DMR, the Langevin equation governing propagation in between attitude measurement updates then reads
\begin{align}\label{eq:LangevinSU2}
\dot{q} = \dfrac{1}{2} \M{\frk{s}}{\omega_m - \eta} q ,
\end{align} 
where $\cal{M}_{\frk{s}}$ is given by \eqref{def:Ms}. Eqn.~\eqref{eq:LangevinSU2} is precisely of the form \eqref{eq:LangevinEquationG}. 

The first matter of business is to compute $\adbar$, defined in \eqref{def:adbar}, for the Lie algebra $\frk{s}$. From the identity \eqref{eq:Omega_Commutator}, it can be seen that $\adbar_{\xi} = - 2 [ \xi ]_{\times}$ for $\xi \in \R^3$. Therefore, $\overline{W}_{\!\!\frk{s}} (\xi) = \left( \int_0^1 e^{- 2 s [\xi]_{\times}} \dd s \right)^{-1}$. To find a closed-form for $\overline{W}_{\!\!\frk{s}} (\xi)$, it is necessary to recall the Rodrigues' rotation formula:
\begin{align}\label{eq:Rodrigues}
e^{s [\xi]_{\times}} = \frac{\xi \xi^T}{\| \xi \|^2} + \cos{(s \| \xi \|)} \left( I_3 - \frac{\xi \xi^T}{\| \xi \|^2} \right) + \frac{\sin{(s \| \xi \|)}}{\| \xi \|} [\xi]_{\times} .
\end{align}
Applying \eqref{eq:Rodrigues} to compute $\int_0^1 e^{- 2 s [\xi]_{\times}} \dd s$, taking the matrix inverse, and simplifying, yields
\begin{align}\label{eq:Wbar_su2}
\overline{W}_{\!\!\frk{s}} (\xi) = \frac{\xi \xi^T}{\| \xi \|^2} + \kappa (\xi) \left( I_3 - \frac{\xi \xi^T}{\| \xi \|^2} \right) + [\xi]_{\times} ,
\end{align}
where $\kappa : B_{\pi} (0) \rightarrow (- \infty , 1]$ is the radial function
\begin{align}\label{def:kappa}
\kappa (\xi) = \| \xi \| \cot{\| \xi \|} .
\end{align}
As a final note, observe that in the $\SU{2}$-case, the ``constant" piece of the Langevin equation $X \equiv \frac{1}{2} \M{\frk{s}}{\omega_m}$, and so $\adbar_{\V{\frk{g}}{X}} \equiv - [\omega_m]_{\times}$. Plugging the previous results into \eqref{eq:TSSDE} yields
\begin{align}\label{eq:TSSDE_su2}
\dot{\xi} = - [ \omega_m ]_{\times} \xi + \frac{1}{2} \overline{W}_{\!\!\frk{s}}(\xi) \eta .
\end{align}

The remainder of this example is devoted to a discussion of the FPE associated with \eqref{eq:TSSDE_su2}, its fundamental solution, and the relevance to optimal attitude filtering. Much of what is covered here is in similar spirit to \cite{Markley_SO3}. In the interest of notational simplicity, for the remainder of this subsection, $\overline{W}_{\!\!\frk{s}}(\xi)$ will be denoted by $G (\xi) \equiv \overline{W}_{\!\!\frk{s}}(\xi)$. For brevity, the interested reader is referred to \cite{do1992riemannian,rosenberg1997laplacian,berger2003panoramic,faraut2008analysis} for additional information regarding the following terminology.

Using the identity $\sum_i \partial_{\xi_i} G_{ij} (\xi) = 2 ( 1 - \kappa (\xi) ) \xi_j / \| \xi \|^2$, the FPE for the transition probability density $p : [0,\infty) \times B_{\pi} (0) \rightarrow [0,\infty)$ associated with \eqref{eq:TSSDE_su2} reads
\begin{align}\label{eq:FPE_su2}
\scalemath{0.95}{
\partial_t p = \diver{ \left( [ \omega_m ]_{\times} \xi + \frac{1 - \kappa (\xi)}{\|\xi\|^2} G (\xi) Q_{\eta} \xi \right) p } + \frac{1 - \kappa (\xi)}{\|\xi\|^2} \xi \cdot Q_{\eta} G^T (\xi) \nabla p + \frac{1}{2} G^T (\xi) \nabla \cdot Q_{\eta} G^T (\xi) \nabla p .
}
\end{align}
The second-order differential operator $G^T (\xi) \nabla \cdot Q_{\eta} G^T (\xi) \nabla$ appearing in \eqref{eq:FPE_su2} is the Laplace-Beltrami operator associated with the Riemannian metric 
\begin{align}\label{eq:Metric_GeneralARW}
\gamma (\xi) = ( G (\xi) Q_{\eta} G^T (\xi) )^{-1} 
\end{align}
on $\S^3$. For $Q_{\eta} = I_3$, $\gamma$ simplifies to
\begin{align}\label{eq:RoundMetric_ExponentialCoordinates}
\gamma (\xi) = \dfrac{\xi \xi^T}{\|\xi\|^2} + h(\xi) \left( I_3 - \dfrac{\xi \xi^T}{\|\xi\|^2} \right) ,
\end{align}
where $h : \R \rightarrow [0,\infty)$ is the radial function
\begin{align}\label{def:h}
h (\xi) = \frac{\sin^2{(\|\xi\|)}}{\|\xi\|^2} .
\end{align}
Furthermore, the FPE \eqref{eq:FPE_su2} simplifies to
\begin{align}\label{eq:FPE_su2_isotropic}
\partial_t p = \diver{ \left( [ \omega_m ]_{\times} \xi + \frac{1 - \kappa (\xi)}{\|\xi\|^2} \xi \right) p } + \frac{1}{2} \left( \frac{h(\xi) - 1}{\|\xi\|^2} \Delta_{\S^2} + \Delta \right) p ,
\end{align}
where $\Delta$ is the standard Laplacian on $\R^3$, and $\Delta_{\S^2}$ is the standard Laplace-Beltrami operator on the unit $2$-sphere. Formula \eqref{eq:RoundMetric_ExponentialCoordinates} is the expression for the standard round metric on $\S^3$ written in exponential coordinates. In general, the metric \eqref{eq:Metric_GeneralARW} is obtained by left-translating the inner product induced by $Q_{\eta}$ on the Lie algebra of $\S^3$. Therefore, the FPE \eqref{eq:FPE_su2} is closely related to the heat equation on $\S^3$ equipped with an arbitrary left-invariant Riemannian metric on $\S^3$.  This is natural considering that the SDE \eqref{eq:TSSDE_su2} was derived from a diffusive process on $\S^3$, i.e. eqn.~\eqref{eq:LangevinSU2}.

Schulman \cite{Schulman68} was the first to write down a closed-form expression for the heat kernel for $\S^3$ equipped with the round metric \eqref{eq:RoundMetric_ExponentialCoordinates}. Leveraging Schulman's formula, it is possible to give the fundamental solution $K : [0,\infty) \times B_{\pi} (0) \rightarrow (0,\infty)$ to \eqref{eq:FPE_su2_isotropic} when $\omega_m = 0$:
\begin{align}\label{eq:FPE_su2_iso_FS}
K (t, \xi) = \frac{h(\xi) e^{t / 2}}{(2 \pi t)^{3/2}} \sum_{n \in \Z} \frac{ \| \xi \| + 2 \pi n }{ \sin{\|\xi\|} } e^{- \frac{ \left( \| \xi \| + 2 \pi n \right)^2 }{ 2 t } } . 
\end{align}
One may verify that $K \geq 0$ and is normalized on $B_{\pi} (0) \subset \R^3$, and hence defines a probability density on $B_{\pi} (0)$. If $p_0 : B_{\pi} (0) \rightarrow [0,\infty)$ is some initial probability density it follows from the left-invariance of the heat kernel \cite{faraut2008analysis} that
\begin{align}\label{eq:Convolution_Soln_su2}
p (t , \xi) = \int_{B_{\pi} (0)} K(t,z(\xi,\delta)) p_0 (\delta) \dd \delta 
\end{align}
is normalized, satisfies \eqref{eq:FPE_su2_isotropic} (with $\omega_m = 0$), and $\lim_{t \rightarrow 0} p(t,\xi) = p_0 (\xi)$, where
\begin{align}\label{def:z}
z(\xi , \delta) = \V{\frk{s}}{\BCH{e^{\M{\frk{s}}{\xi}}}{e^{-\M{\frk{s}}{\delta}}}} 
\end{align}
Formula \eqref{eq:Convolution_Soln_su2} essentially reflects the fact that convolution against the heat kernel on a Lie group yields the propagated solution to the heat equation on the group. Convolution is in terms of the group operation, which in this case arises from the underlying group structure on $\S^3$. From the Chapman-Kolmogorov equation, \eqref{eq:FPE_su2_iso_FS} essentially gives the class of probability distributions which are invariant under propagation through \eqref{eq:TSSDE_su2} in the case of isotropic $Q_{\eta}$. This observation should be compared with the canonical example of the heat kernel on $\R^n$ endowed with the standard Euclidean metric. This is the Gaussian function $K(x,y,t) = \frac{1}{(4 \pi t)^{d/2}} \exp{\left( - \frac{\| x - y \|^2}{4t} \right)}$, for $x,y \in \R^n$, which satisfies the heat equation $\partial_t K = \Delta K$ on $\R^n$. The Chapman-Kolmogorov equation for the previous heat kernel encodes the well-known fact that the convolution of two Gaussians is again a Gaussian, a fact which is the essence of the propagation step in the classical linear Kalman filter. Notice that \eqref{eq:FPE_su2_iso_FS} is well-approximated by the Gaussian $e^{- \|\xi\|^2 / 2t}/(2 \pi t)^{3/2}$ for $t \ll 1$ and $\| \xi \| \ll \pi$, which provides theoretical support for earlier claims that CGs are the appropriate class of distributions to filter with respect to on Lie groups, even in the presence of noise.

Using left-invariance, a relatively simple modification of \eqref{eq:FPE_su2_iso_FS} yields the fundamental solution, $K_{\omega_m} (t, \xi)$, for \eqref{eq:FPE_su2_isotropic} when $\omega_m \neq 0$: $K_{\omega_m} (t, \xi) = K(t,z(\xi,\omega_m))$, where $z$ was defined in \eqref{def:z}. A further generalization of \eqref{eq:FPE_su2_iso_FS} to the case of non-isotropic $Q_{\eta}$ (i.e., the fundamental solution to \eqref{eq:FPE_su2}) would require, at a minimum, a closed-form for the geodesic distance on $\S^3$ equipped with an arbitrary left-invariant Riemannian metric, and such a formula appears to be unknown. A class of likelihood functions which leave \eqref{eq:FPE_su2_iso_FS} invariant under a Bayesian update is also unknown, and we suspect there is no such "natural" observation process (as compared to the classical linear Kalman filter where linear measurements with Gaussian distributed measurement noise leaves the class of Gaussian priors invariant under a Bayesian update). If these problems could be solved, however, then one would have a theory of attitude filtering on $\S^3$ under the paradigm of DMR that is analogous to Kalman's classical theory on Euclidean space. 

As described at the end of Section \ref{sub:propagation}, there is a short-time expansion that approximates \eqref{eq:FPE_su2_isotropic} to yield a simpler FPE for which the moment evolution equations may be solved exactly. This approximate FPE is obtained by truncating the Taylor expansion of $g$ \eqref{def:kappa} and $h$ \eqref{def:h} to second-order, and then substituting these truncated Taylor expansions for $g$ and $h$ in \eqref{eq:FPE_su2_isotropic}. The resulting FPE reads
\begin{align}\label{eq:FPE_su2_isotropic_simplified}
\partial_t p = \diver{ \left( [ \omega_m ]_{\times} + \frac{q_{\eta}^2}{12} I_3 \right) \xi p } + \frac{q_{\eta}^2}{2} \left( \frac{1}{12} \Delta_{\S^2} + \Delta \right) p ,
\end{align}
where $q_{\!\eta}$ is defined via $Q_{\eta} = q_{\eta}^2 I_3$. The moment evolution equations for the density which satisfies \eqref{eq:FPE_su2_isotropic_simplified} read
\begin{align}\label{eq:MomentEvolution_su2_simplified}
\left\lbrace \begin{array}{l}
\dfrac{\dd \wh{\xi}}{\dd t} = - \left( [\omega_m]_{\times} + \dfrac{q_{\!\eta}^2}{6} I_3 \right) \wh{\xi} \\[5pt] 
\dfrac{\dd P}{\dd t} = - \left( [\omega_m]_{\times} + \dfrac{q_{\!\eta}^2}{12} I_3 \right) P - P \left( [\omega_m]_{\times} + \dfrac{q_{\!\eta}^2}{12} I_3 \right)^T - \dfrac{q_{\!\eta}^2}{12} \left( P - \tr{P} I_3 \right) + q_{\!\eta}^2 I_3 ,
\end{array} \right.
\end{align}
where $\wh{\xi} := \E{\xi}$ and $P = \Cov{\xi}$. The evolution equations \eqref{eq:MomentEvolution_su2_simplified} are both \textit{linear}, a desirable feature if computational speed up is required. 

\subsection{$\SE{3}$}\label{sub:SE3}

Consider the special Euclidean group in $3$ dimensions, $\SE{3}$, which is the semidirect product of $\SO{3}$ and $\R^3$, typically written as $\SE{3} = \SO{3} \ltimes \R^3$. An element $A \in \SE{3}$ will be represented as a $4 \times 4$ matrix of the form
\begin{align}\label{eq:SE3_groupelement}
A = \left( \begin{array}{cc}
R & r \\
0 & 1
\end{array} \right) \in \GL{4} ,
\end{align}
where $R \in \SO{3}$ and $r \in \R^3$. A physical context to the elements of $\SE{3}$ may be ascribed as the position $r \in \R^3$ and attitude $R \in \SO{3}$ of a rigid body. If the rigid body is subject to an angular velocity $\omega \in \R^3$ and $r \in \R^3$ describes its position in the reference frame of the rigid body, then the attitude $R(t)$ and position $r(t)$ of the body at any instance in time $t$ satisfy
\begin{align}\label{eq:SE3_ODEs}
\left\lbrace \begin{array}{lc}
\dot{R} (t) = [\omega]_{\times} R (t) \\[5pt]
\dot{r} (t) = [\omega]_{\times} r(t) + v .
\end{array} \right.
\end{align}

The ODEs \eqref{eq:SE3_ODEs} may be related to the INS problem as follows. Let $R \equiv T_i^b$ be the rotation from an inertial frame to a body frame. Then,
\begin{align}
\dot{R} = \dot{T}_i^b = T_i^b [\omega^i_{bi}]_{\times} = T_i^b [\omega^i_{bi}]_{\times} T_b^i T_i^b = [\omega^b_{bi}]_{\times} T_i^b ,
\end{align}
where $\omega^i_{ib}$ denotes the angular rate of the inertial frame with respect to the body frame, expressed in the inertial frame, and likewise for $\omega^b_{bi}$. Let $r \equiv r^b$, where $r^b = T_i^b r^i$, i.e. the position expressed in the body frame. Then,
\begin{align}
\dot{r} = \dot{r}^b = T_i^b \left( [\omega^i_{bi}]_{\times} r^i + \dot{r}^i \right) = T_i^b [\omega^i_{bi}]_{\times} T_b^i r^b + v^b = [\omega^b_{bi}]_{\times} r^b + v^b ,
\end{align}
where $v^b = T_i^b v^i$ is the velocity expressed in the body frame. From the previous two equations, $\omega \equiv \omega^b_{bi}$ and $v \equiv v^b$ in \eqref{eq:SE3_ODEs}. 

The Lie algebra $\se{3}$ of $\SE{3}$ consists of matrices of the form
\begin{align}\label{eq:SE3_LieAlgera}
\M{\se{3}}{\omega , v} = \left( \begin{array}{cc}
[\omega]_{\times} & v \\
0 & 0
\end{array} \right) ,
\end{align}
where $\omega , v \in \R^3$. The exponential map is
\begin{align}\label{eq:SE3_ExponentialMap}
e^{\M{\se{3}}{\omega , v}} = \left( \begin{array}{cc}
e^{[\omega]_{\times}} & \overline{J}_{\!\so{3}} (-\omega) v \\
0 & 1
\end{array} \right) ,
\end{align}
where
\begin{align}\label{eq:Jbar_so3}
\overline{J}_{\!\so{3}} (\omega) := \int_0^1 e^{-s [\omega]_{\times}} \dd s . 
\end{align}
Therefore, the curve
\begin{align}
A(t) = e^{t \M{\se{3}}{\omega , v}} A(0) = \left( \begin{array}{cc}
e^{t [\omega]_{\times}} R_0 & e^{t[\omega]_{\times}} r_0 + \int_0^t e^{(t-s) [\omega]_{\times}} v \dd s \\
0 & 1
\end{array} \right) \in \SE{3} ,
\end{align}
is the solution of \eqref{eq:SE3_ODEs} written in a Lie algebraic way. The vector field in $\dot{A} (t) = \M{\se{3}}{\omega , v} A (t)$ is of the form \eqref{eq:GroupAffineCharacterization} (with $D = 0$) and is therefore group-affine. Hence, in light of the group structure of $\SE{3}$, the "natural" state vector to estimate is $x = (T_i^b , r^b)$, assuming one desires to use CGs over $\SE{3}$ to capture the uncertainty associated with $x$. 

Similar to the $\SU{2}$-case, the Lie algebra coordinates $(\omega , v) \in \R^6$ are treated as the measured quantities. The measured angular rate and velocity satisfy $\omega_m = \omega + \eta$ and $v_m = v + \zeta$, respectively, with $\eta \sim \cal{N}^3 (0 , Q_{\eta})$ and $\zeta \sim \cal{N}^3 (0 , Q_{\zeta})$. Under DMR, the Langevin equation governing propagation in between measurement updates then reads
\begin{align}\label{eq:LangevinSE3}
\dot{A} = \left( \M{\se{3}}{\omega_m , v_m} - \M{\se{3}}{\eta , \zeta} \right) A . 
\end{align} 
Eqn.~\eqref{eq:LangevinSE3} is of the form \eqref{eq:LangevinEquationG}. Again, the first task is to compute $\adbar$ for $\se{3}$. A computation shows, for $(\omega , v) \in \R^3 \times \R^3$,
\begin{align}\label{eq:adbarSE3}
\adbar_{(\omega , v)} = \left( \begin{array}{cc}
[\omega]_{\times} & 0 \\
~[v]_{\times} & [\omega]_{\times} 
\end{array} \right) . 
\end{align}
Using the Van Loan method \cite{van1978computing} to compute the matrix exponential yields
\begin{align}\label{eq:WbarSE3}
\overline{W}_{\!\!\se{3}}(\omega, v) = \left( \begin{array}{cc}
\overline{W}_{\!\!\so{3}} (\omega) & 0 \\
- \overline{W}_{\!\!\so{3}} (\omega) N (\omega, v) \overline{W}_{\!\!\so{3}} (\omega) & \overline{W}_{\!\!\so{3}} (\omega)
\end{array} \right) ,
\end{align}
where
\begin{align}\label{def:WbarSO3}
\overline{W}_{\!\!\so{3}} (\omega) := \left( \int_0^1 e^{s [\omega]_{\times}} \dd s \right)^{-1} ,
\end{align}
and
\begin{align}\label{def:N_SE3}
N (\omega , v) := \int_0^1 e^{t [\omega]_{\times}} [ \int_0^t e^{-s [\omega]_{\times}} v \dd s ]_{\times} \dd t .
\end{align}
Note that $\overline{W}_{\!\!\so{3}} (\omega)$ is essentially given by \eqref{eq:Wbar_su2} after a simple rescaling of variables. Plugging \eqref{eq:adbarSE3}-\eqref{eq:WbarSE3} into \eqref{eq:TSSDE} yields the TSSDE for $\SE{3}$ corresponding to \eqref{eq:LangevinSE3}:
\begin{align}\label{eq:SE3_TSSDE}
\frac{\dd}{\dd t} \left( \begin{array}{c}
\delta \\
w
\end{array} \right) = \left( \begin{array}{cc}
[\omega_m]_{\times} & 0 \\
~[v_m]_{\times} & [\omega_m]_{\times} 
\end{array} \right) \left( \begin{array}{c}
\delta \\
w
\end{array} \right) + \overline{W}_{\!\!\se{3}}(\delta , w) \left( \begin{array}{c}
\eta \\
\zeta
\end{array} \right) 
\end{align}
in which $\delta , w \in \R^3$ are the coordinates parameterizing the Lie algebra $\se{3}$. 


\subsection{$\SE{2,3}$}\label{sub:SE23}

Ultimately, $\SE{3}$ is \textit{not} the Lie group appropriate to describe state estimation from IMU measurements, since it is not the body frame velocity $v^b$ the IMU measures, but rather the body frame acceleration $a^b$ (minus gravity). This complicates things in an interesting way. First, an appropriate Lie group needs to be identified. Inspired by the results of Section \ref{sub:SE3}, consider the state vector given by $x = (T_i^b , r^b , v^b)$. This state vector form motivates the group $\SE{2,3} \subset \GL{5}$ given by matrices of the form 
\begin{align}
A = \left( \begin{array}{ccc}
R & v & r \\
0 & 1 & 0 \\
0 & 0 & 1
\end{array} \right) ,
\end{align}
where $R \in \SO{3}$ and $r , v \in \R^3$ (thought of as a $3 \times 1$ column vectors). This makes $\SE{2,3} = \SO{3} \ltimes \R^6$. To the best of our knowledge, the group $\SE{2,3}$ first appeared in \cite{barrau2016invariant}, and is referred to as the group of \textit{double direct spatial isometries}.

Notice that $\dot{v}^b = [\omega^b_{bi}]_{\times} v^b + a^b + T_i^b g^i$, where $g^i \in \R^3$ is the gravitational acceleration in an inertial frame and will be treated as constant (i.e., independent of $r^i$). (In what follows, drop the superscripts $b$ and $i$ for notational simplicity.) The equations of motion on $\SE{2,3}$ read
\begin{align}\label{eq:SE23_ODEs}
\left\lbrace \begin{array}{lc}
\dot{R} (t) = [\omega]_{\times} R(t) \\[5pt]
\dot{r} (t) = [\omega]_{\times} r(t) + v(t) \\[5pt]
\dot{v} (t) = [\omega]_{\times} v(t) + a + R (t) g ,
\end{array} \right.
\end{align}
or, equivalently,
\begin{align}\label{eq:DynamicsSE23}
A' (t) = \M{\se{2,3}}{\omega , a + R (t) g , v(t)} A (t) ,
\end{align}
where 
\begin{align} \label{eq:SE23Xdef}
\M{\se{2,3}}{\omega , a , v} = \left( \begin{array}{ccc}
[\omega]_{\times} & a & v \\
0 & 0 & 0 \\
0 & 0 & 0
\end{array} \right) ,
\end{align}
is a member of the Lie algebra $\se{2,3}$ of $\SE{2,3}$. 

Eqn.~\eqref{eq:DynamicsSE23} doesn't fall into the class of constant dynamical systems because of the terms involving the gravitational acceleration $g$ and the velocity variables $v$. This results in a distinct departure from the $\SU{2}$ and $\SE{3}$ examples considered so-far. In an example INS application under DMR (see Sec.~\ref{sec:AttEst}), $\omega$ and $a$ in \eqref{eq:SE23_ODEs} are replaced by $\omega_m - \eta$ and $a_m - \zeta$, where $\omega_m$ and $a_m$ are the measured angular rate and body frame acceleration, respectively, and $\eta \sim \cal{N}^3 (0,Q_{\eta})$ and $\zeta \sim \cal{N}^3 (0,Q_{\zeta})$. $\omega_m$ and $a_m$ are treated as constants over the time interval of integration. A model for $g$ (in the inertial frame) is typically assumed. The vector field $A \mapsto \M{\se{2,3}}{\omega_m , a_m + R g , v} A$ is verified to be group-affine, \textit{assuming} $g$ independent of $r$. The fidelity of this assumption depends on the scenario being considered.  According to Theorem \ref{thm:Characterization}, there must exist a linear equation of the form \eqref{eq:TSSDE_2} on the Lie algebra corresponding to the ODE $\dot{A} = \M{\se{2,3}}{\omega_m , a_m + R g , v} A$. This will be shown to be the case. It is noteworthy that the DP group law on $\SO{3} \times \R^6$, corresponding to the state vector $x = (T_i^b , r^i , v^i)$, fails to make the dynamics
\begin{align*}
\left\lbrace \begin{array}{lc}
\dot{R} (t) = [\omega]_{\times} R(t) \\[5pt]
\dot{r} (t) = v(t) \\[5pt]
\dot{v} (t) = R^T (t) a + g 
\end{array} \right.
\end{align*}
group-affine. 

Let the Lie algebra coordinates be partitioned as $\xi = (\delta , u , w) \in \R^9$. The exponential map on $\SE{2,3}$ reads
\begin{align}\label{eq:SE23_ExponentialMap}
e^{\M{\se{2,3}}{\xi}} = \left( \begin{array}{ccc}
e^{[\delta]_{\times}} & \overline{J}_{\!\so{3}} (-\delta) u & \overline{J}_{\!\so{3}} (-\delta) w \\
0 & 1 & 0 \\
0 & 0 & 1
\end{array} \right) ,
\end{align}
where $\overline{J}_{\!\so{3}}$ was introduced in \eqref{eq:Jbar_so3}. Supply \eqref{eq:DynamicsSE23} with the initial condition $A(0) \equiv (R_0 , v_0 , r_0)$. The MAP $\mu (t) \in \SE{2,3}$ is then given by
\begin{align}\label{eq:SE23_MAP}
\mu (t) = \left( \begin{array}{ccc}
e^{t [ \omega_m ]_{\times}} R_0 & \wh{v} (t) & e^{t [ \omega_m ]_{\times}} r_0 + \int_0^t e^{(t-s) [ \omega_m ]_{\times}} \wh{v} (s) \dd s \\
0 & 1 & 0 \\
0 & 0 & 1
\end{array} \right) ,
\end{align}
where $\wh{v} (t) = e^{t [ \omega_m ]_{\times}} v_0 + \int_0^t e^{(t-s) [ \omega_m ]_{\times}} ( a_m + e^{s [ \omega_m ]_{\times}} R_0 g ) \dd s$. Starting from eqn.~\eqref{eq:ODE_xi_groupaffine}, a derivation using \eqref{eq:SE23_ExponentialMap}-\eqref{eq:SE23_MAP} shows that eqn.~\eqref{eq:TSSDE_2} in the case of $\SE{2,3}$ becomes
\begin{align}\label{eq:SE23_TSSDE}
\frac{\dd}{\dd t} \left( \begin{array}{c}
\delta \\
u \\
w 
\end{array} \right) = \left( \begin{array}{ccc}
~[\omega_m]_{\times} & 0 & 0 \\
~[a_m]_{\times} & [\omega_m]_{\times} & 0 \\
0 & I & [\omega_m]_{\times} 
\end{array} \right) \left( \begin{array}{c}
\delta \\
u \\
w 
\end{array} \right) - \overline{W}_{\!\!\se{2,3}} (\xi) \left( \begin{array}{c}
\eta \\
\zeta \\
0 
\end{array} \right) ,
\end{align}
where
\begin{align}\label{eq:Wbar_se23}
\overline{W}_{\!\!\se{2,3}} (\xi) = \left( \begin{array}{ccc}
\overline{W}_{\!\!\so{3}} (\delta) & 0 & 0 \\
- \overline{W}_{\!\!\so{3}} (\delta) N(\delta , u) \overline{W}_{\!\!\so{3}} (\delta) & \overline{W}_{\!\!\so{3}} (\delta) & 0 \\
- \overline{W}_{\!\!\so{3}} (\delta) N(\delta , w) \overline{W}_{\!\!\so{3}} (\delta) & 0 & \overline{W}_{\!\!\so{3}} (\delta)
\end{array} \right)  ,
\end{align}
with $N : \R^6 \rightarrow \R^{3 \times 3}$ given by \eqref{def:N_SE3}. It is interesting that, even in the noiseless case, the evolution of the MAP is not "constant", but the evolution on the Lie algebra \textit{is} linear. It is also noteworthy that the gravitational acceleration does not appear in \eqref{eq:SE23_TSSDE}. This provides a non-trivial example in which filtering on the group with a CG yields a propagation stage with complexity equal to that of standard Kalman filtering, but which \emph{exactly} incorporates nonlinearity missed by conventional Kalman approaches.

\section{Gyro Bias Estimation}{\label{sec:GyroBias}

This section addresses the problem of jointly estimating both the attitude of a rigid body and the bias contributing to the measured angular rate from a gyroscope, as in the Farrenkopf model \eqref{eq:Farrenkopf}. The underlying state space here will be the $6$ dimensional manifold $\SO{3} \times \R^3$, where the state to-be-estimated is $(R , \beta)$, with $R$ be the rotation matrix representing the attitude of the rigid body and $\beta$ representing the gyro bias. The Langevin equation describing the evolution of the state $(R,\beta)$ reads
\begin{align}\label{eq:LangevinGyroBias}
\left\lbrace \begin{array}{l}
\dot{R} = [\omega_m - \beta - \eta]_{\times} R \\[5pt]
\dot{\beta} = \zeta ,
\end{array} \right.
\end{align}
where $\eta , \zeta$ are defined in \eqref{eq:Farrenkopf}. 

There is more than one group law one may assign to pairs $(R , \beta)$ which turns $\SO{3} \times \R^3$ into a Lie group. The simplest is the DP group law, in which two pairs $(R_1 , \beta_1) , (R_2 , \beta_2)$ are combined according to $(R_1 , \beta_1)(R_2 , \beta_2) = (R_1 R_2 , \beta_1 + \beta_2)$. Another is the $\SE{3}$ group law (i.e., a semidirect product), where $(R_1 , \beta_1)(R_2 , \beta_2) = (R_1 R_2 , \beta_1 + R_1 \beta_2)$. Unfortunately, neither of these groups make the vector field $f (R, \beta) = ( [\omega_m - \beta]_{\times} R , 0 )$ group-affine. However, a time-parameterized family of group operations on $\SO{3} \times \R^3$ is identified that ensures the vector field $f$ is "almost" group-affine (and is group-affine when restricted to $\SO{3}$). This family corresponds to the $\SE{3}$ group law to zeroth order in time, thus providing theoretical motivation for this choice of group law over, e.g., the DP group law. Furthermore, the numerical superiority of the $\SE{3}$ group law over the DP group law for this state estimation problem is demonstrated in Section \ref{sub:experiment}.

Identify $\SO{3} \times \R^3$ with $\SO{3} \times \so{3}$ via the map $(R , \beta) \mapsto (R , [\beta]_{\times})$. Consider a family of Lie groups $G_t$ parameterized by time $t \in \R$ where for each $t$ the underlying manifold of $G_t$ is simply $\SO{3} \times \so{3}$. The group operation is a function of $t$, and is defined via
\begin{align}\label{eq:timedependentGL}
\left( \begin{array}{c}
R_1 \\
~[\beta_1]_{\times}
\end{array} \right) \circ_t \left( \begin{array}{c}
R_2 \\
~[\beta_2]_{\times}
\end{array} \right) = \left( \begin{array}{c}
R_1 R_2 \\
t^{-1} \BCH{ t R_1 [\beta_2]_{\times} R_1^T }{t [\beta_1]_{\times}}
\end{array} \right) .
\end{align}
It may verified that $\circ_t$ yields a genuine group law for each $t \neq 0$. (The only nontrivial aspect of this is checking the associativity of $\circ_t$. This fact essentially boils down to the identity $\BCH{\BCH{X}{Y}}{Z} = \BCH{X}{\BCH{Y}{Z}}$ for $X,Y,Z \in \frk{g}$, which, in turn, follows from associativity of matrix multiplication.) From the BCH series \eqref{eq:BCH_series}, observe that
\begin{align}
\frac{1}{t} \BCH{ t R_1 [\beta_2]_{\times} R_1^T }{t [\beta_1]_{\times}} = [\beta_1]_{\times} + R_1 [\beta_2]_{\times} R_1^T - \frac{t}{2} [ [\beta_1]_{\times} , R_1 [\beta_2]_{\times} R_1^T ] + \mathcal{O} (t^2) . 
\end{align}
From the identity $R_1 [\beta_2]_{\times} R_1^T = [R_1 \beta_2]_{\times}$, the group law $\circ_t$ is seen to be equivalent to the $\SE{3}$ group law at $t = 0$. Finally, we note that there is an explicit representation of the group law $\circ_t$ on $\SO{3} \times \R^3$ that doesn't pass through $\R^3 \simeq \so{3}$ and utilizes the closed-form for the BCH series on $\so{3}$. Indeed, let $c \in \R^3$ be the vector such that $[c]_{\times} = \BCH{[a]_{\times}}{[b]_{\times}}$. Then, $c$ satisfies $\|c\| = \arccos{ \left( \cos{ \left( \| a \| \right)} \cos{ \left( \| b \| \right)} - \sin{ \left( \| a \| \right)} \sin{ \left( \| b \| \right)} \frac{\langle a , b \rangle}{\| a \| \| b \|} \right) }$ and
\begin{align*}
 \frac{c}{\|c\| \csc{(\|c\|)}} = \frac{ \sin{\left(\| a \|\right)} \cos{\left(\| b \|\right)}}{\|a\|} a + \frac{ \sin{\left(\| b \|\right)} \cos{\left(\| a \|\right)}}{\|b\|} b - \frac{ \sin{\left(\| a \|\right)} \sin{\left(\| b \|\right)} }{ \| a \| \| b \| } a \times b .
\end{align*}

For simplicity, set $\omega_m = 0$. Let us denote by $\psi_t : \SO{3} \times \so{3} \rightarrow \SO{3} \times \so{3}$ the flow associated with \eqref{eq:LangevinGyroBias} in the absence of noise. In particular,
\begin{align}\label{eq:Flow_Determinisitc_GyroBias}
\psi_t (R , [\beta]_{\times}) = \left( \begin{array}{c}
e^{- t [\beta]_{\times}} R \\
~[\beta]_{\times}
\end{array} \right) . 
\end{align}
From \cite[Proposition 8]{barrau2019linear}, group-affinity (under the group law \eqref{eq:timedependentGL}) would be equivalent to the existence of $\varphi_t \in \Aut{G_t}$ and $\nu_t \in G_t$ such that the flow $\psi_t$ \eqref{eq:Flow_Determinisitc_GyroBias} satisfies $\psi_t (R , [\beta]_{\times}) = \varphi_t (R , [\beta]_{\times}) \circ_t \nu_t$. Since an automorphism maps the identity to itself, it must be the case that $\nu_t = (I_3,0)$. In other words, the flow $\psi_t$ must be an automorphism of $G_t$, if the associated vector field is group-affine. Let us now analyze how close $\psi_t$ in \eqref{eq:Flow_Determinisitc_GyroBias} is to being an automorphism.

The defining property of an automorphism asserts that
\begin{align}\label{eq:AutomorphismIdentity}
\psi_t ( R_1 R_2 , t^{-1} \BCH{ t R_1 [\beta_2]_{\times} R_1^T }{ t [\beta_1]_{\times} } ) = \psi_t (R_1, \beta_1) \circ_t \psi_t (R_2 , \beta_2) ,
\end{align}
for all $(R_1 , [\beta_1]_{\times}) , (R_2 , [\beta_2]_{\times}) \in G_t$. For the LHS of \eqref{eq:AutomorphismIdentity}:
\begin{align}\label{eq:AutomorphismIdentity_LHS}
\psi_t ( R_1 R_2 , t^{-1} \BCH{ t R_1 [\beta_2]_{\times} R_1^T }{ t [\beta_1]_{\times} } ) = \left( \begin{array}{c}
e^{- \BCH{ t R_1 [\beta_2]_{\times} R_1^T }{ t [\beta_1]_{\times} } } R_1 R_2 \\
t^{-1} \BCH{ t R_1 [\beta_2]_{\times} R_1^T }{ t [\beta_1]_{\times} }
\end{array} \right) ,
\end{align}
while the RHS of \eqref{eq:AutomorphismIdentity} reads
\begin{align}\label{eq:AutomorphismIdentity_RHS}
\psi_t (R_1, \beta_1) \circ_t \psi_t (R_2 , \beta_2) = \left( \begin{array}{c}
e^{- t [\beta_1]_{\times}} R_1 e^{-t [\beta_2]_{\times}} R_2 \\
t^{-1} \BCH{t e^{- t [\beta_1]_{\times}} R_1 [ \beta_2 ]_{\times} R_1^T e^{t [\beta_1]_{\times}}}{t [\beta_1]_{\times}}
\end{array} \right) . 
\end{align}
Analyzing \eqref{eq:AutomorphismIdentity_LHS}-\eqref{eq:AutomorphismIdentity_RHS}, the bias terms differ in the first argument of BCH by a conjugation by $e^{- t [\beta_1]_{\times}}$. For the $\SO{3}$ term in \eqref{eq:AutomorphismIdentity_RHS}, notice that
\begin{align*}
e^{- t [\beta_1]_{\times}} R_1 e^{-t [\beta_2]_{\times}} R_2 = e^{- t [\beta_1]_{\times}} e^{- t R_1 [\beta_2]_{\times} R_1^T} R_1 R_2 = e^{\BCH{ - t [\beta_1]_{\times}}{- t R_1 [\beta_2]_{\times} R_1^T}} R_1 R_2 . 
\end{align*}
Recalling the BCH symmetry $\BCH{X}{Y} = - \BCH{-Y}{-X}$, one concludes that
\begin{align*}
e^{- t [\beta_1]_{\times}} R_1 e^{-t [\beta_2]_{\times}} R_2 = e^{- \BCH{ t R_1 [\beta_2]_{\times} R_1^T }{ t [\beta_1]_{\times} } } R_1 R_2 .
\end{align*}
Hence, the $\SO{3}$ pieces of \eqref{eq:AutomorphismIdentity_LHS} and \eqref{eq:AutomorphismIdentity_RHS} agree, ensuring that $\psi_t$ is an automorphism of $G_t$ when restricted to $\SO{3}$. A modification of the group law \eqref{eq:timedependentGL} which makes the bias terms in \eqref{eq:AutomorphismIdentity_LHS}-\eqref{eq:AutomorphismIdentity_RHS} agree while simultaneously making the $\SO{3}$ terms agree has not been identified.

The question that now must be addressed is the derivation of SDEs on the Lie algebra corresponding to the Langevin equations \eqref{eq:LangevinGyroBias}, for both the $\SE{3}$ and DP group laws on $\SO{3} \times \R^3$. Since neither of these group laws make the deterministic part of \eqref{eq:LangevinGyroBias} group-affine, the TSSDE \eqref{eq:TSSDE_2} derived in Section \ref{sec:TSSDE} does not directly apply. Instead, one must return to eqn.~\eqref{eq:ODE_xi}, which was derived without making any assumptions on $f$. The computation of the terms in \eqref{eq:ODE_xi} will be carried out for the $\SE{3}$ group law on $\SO{3} \times \R^3$, while the DP group law will follow in a similar fashion.

The vector field $f : \SE{3} \rightarrow \se{3}$ corresponding to the deterministic piece of \eqref{eq:LangevinGyroBias} reads
\begin{align}\label{eq:SE3_GyroBias_VF}
f (A) = \M{\se{3}}{\omega_m - \beta , - [\omega_m]_{\times} \beta} A , 
\end{align}
where $\cal{M}_{\se{3}}$ is given by \eqref{eq:SE3_LieAlgera} and $A$ is given by
\begin{align*}
A = \left( \begin{array}{cc}
R & \beta \\
0 & 1
\end{array} \right) .
\end{align*}
Suppose $g : [0 , \infty) \rightarrow \SE{3}$ satisfies \eqref{eq:LangevinGyroBias} \textit{without noise} and where $g(0) = e^{\M{\se{3}}{\xi_0}} \mu_0$ with $\xi_0 \sim \cal{N}^6 (0 , \Sigma_0)$ and
\begin{align*}
\mu_0 = \left( \begin{array}{cc}
R_0 & \wh{\beta}_0 \\
0 & 1
\end{array} \right)
\end{align*}
is given. Further suppose $g (t)$ for $t > 0$ may be represented as $g (t) = e^{\M{\se{3}}{\xi(t)}} \mu(t)$, where $\mu : [0, \infty) \rightarrow \SE{3}$ equals $\mu (t) = e^{t \M{\se{3}}{\omega_m - \wh{\beta}_0 , -[\omega_m]_{\times} \wh{\beta}_0} } \mu_0$. In other words, $\mu$ follows the same evolution equation as $g$, except the initial condition is known and given by $\mu_0 \in \SE{3}$. From \eqref{eq:ODE_xi}, $\M{\se{3}}{\xi}$ then satisfies the ODE 
\begin{align}\label{eq:TangentSpaceODE_Nonlinear_SE3_GyroBias}
J_{\se{3}} (\M{\se{3}}{\xi}) \M{\se{3}}{\dot{\xi}} \mu = e^{-\M{\se{3}}{\xi}} f (e^{\M{\se{3}}{\xi}} \mu) - f (\mu) ,
\end{align}

Let the Lie algebra coordinates be partitioned as $\xi = (\delta , u) \in \R^3 \times \R^3$. Using \eqref{eq:SE3_GyroBias_VF}, the $\SE{3}$ exponential map \eqref{eq:SE3_ExponentialMap}, and the braiding identity \eqref{eq:braiding}, the RHS of \eqref{eq:TangentSpaceODE_Nonlinear_SE3_GyroBias} may be simplified to read
\begin{align*}
J_{\se{3}} (\M{\se{3}}{\xi}) \M{\se{3}}{\dot{\xi}} & = \left( e^{-\ad_{\M{\se{3}}{\xi}}} - I \right) \M{\se{3}}{\omega_m - \wh{\beta}_0 , - [\omega_m]_{\times} \wh{\beta}_0 }  \\
& \hspace{5mm} + e^{-\ad_{\M{\se{3}}{\xi}}} \M{\se{3}}{\wh{\beta}_0  - B (\delta , u , \wh{\beta}_0) , [\omega_m]_{\times} \left( \wh{\beta}_0  - B (\delta , u , \wh{\beta}_0) \right) } ,
\end{align*}
where
\begin{align}\label{def:Beta}
\tilde{\beta} (\delta , u , \wh{\beta}_0) = e^{[\delta]_{\times}} ( \wh{\beta}_0 + \overline{J}_{\!\so{3}} (\delta) u ) ,
\end{align}
with $\overline{J}_{\!\so{3}}$ given by \eqref{eq:Jbar_so3}. Vectorizing the previous equation, while noting that $\overline{J}_{\!\se{3}}^{-1} (\xi) \left( e^{-\adbar_{\xi}} - I \right) = - \adbar_{\xi}$ and $\overline{J}_{\!\se{3}}^{-1} (\xi) e^{-\adbar_{\xi}} = \overline{W}_{\!\!\se{3}} (\xi)$, where $\overline{W}_{\!\!\se{3}}$ is given by \eqref{eq:WbarSE3}, results in 
\begin{align}\label{eq:TangentSpaceODE_Nonlinear_SE3_GyroBias2}
 \dot{\xi} = - \adbar_{\xi} \left( \begin{array}{c}
 \omega_m - \wh{\beta}_0 \\
 - [\omega_m]_{\times} \wh{\beta}_0
\end{array} \right) + \overline{W}_{\!\!\se{3}} (\xi) \left( \begin{array}{c}
 \wh{\beta}_0 - \tilde{\beta} (\delta , u , \wh{\beta}_0) \\
~ [\omega_m]_{\times} \left( \wh{\beta}_0  - \tilde{\beta} (\delta , u , \wh{\beta}_0) \right)
\end{array} \right) . 
\end{align}
The term $\wh{\beta}_0 - \tilde{\beta} (\delta , u , \wh{\beta}_0)$ represents the deviation that fails to make \eqref{eq:TangentSpaceODE_Nonlinear_SE3_GyroBias2} a linear equation on the Lie algebra. This term may be interpreted as the difference between the gyro bias MAP $\wh{\beta}_0$ and its random fluctuation $\tilde{\beta} (\delta , u , \wh{\beta}_0)$.

Eqn.~\eqref{eq:TangentSpaceODE_Nonlinear_SE3_GyroBias2} is the desired ODE on the Lie algebra. Simplifying this equation further results in valuable insights. Using the formula \eqref{eq:adbarSE3} for $\adbar$ on $\se{3}$ and the explicit formula for $\overline{W}_{\!\!\se{3}}$ in \eqref{eq:WbarSE3}, eqn.~\eqref{eq:TangentSpaceODE_Nonlinear_SE3_GyroBias2} may be simplified to yield the following set of equations for the evolution of $\xi = (\delta , u)$:
\begin{align}\label{eq:TSODE_SE3_GyroBias}
\scalemath{0.95}{
\left\lbrace \begin{array}{l}
\dot{\delta} = [\omega_m]_{\times} \delta - u \\[3pt]
\dot{u} = - [[\omega_m]_{\times} \wh{\beta}_0]_{\times} \delta + [\omega_m - \wh{\beta}_0]_{\times} u - \overline{W}_{\!\!\so{3}} (\delta) \left( N(\delta , u) \overline{W}_{\!\!\so{3}} (\delta) - [\omega_m]_{\times} \right) \left( \wh{\beta}_0  - \tilde{\beta} (\delta , u , \wh{\beta}_0) \right)  ,
\end{array} \right.
}
\end{align}
where $N$ is defined in \eqref{def:N_SE3} and $\overline{W}_{\!\!\so{3}}$ is given by \eqref{def:WbarSO3}. Observe that the equation for $\delta$ in \eqref{eq:TSODE_SE3_GyroBias} is linear, while the equation for $u$ is nonlinear. This is also consistent with the observation that the vector field in \eqref{eq:LangevinGyroBias} is group-affine for $\SE{3}$ (to zeroth-order in $t$) for the attitude piece, but not the bias. Moreover, the second equation for $\dot{u}$ in \eqref{eq:TSODE_SE3_GyroBias} notably depends on the MAP through $\wh{\beta}_0$. This is consistent the observation that, in the course of computing \eqref{eq:ODE_xi_groupaffine}, the MAP was only eliminated from \eqref{eq:ODE_xi} because of the group-affine assumption on $f$. Since the $\SE{3}$ group structure doesn't make the vector field in \eqref{eq:LangevinGyroBias} group-affine, this is expected.

The derivation of eqn.~\eqref{eq:TSODE_SE3_GyroBias} did not include the noise in \eqref{eq:LangevinGyroBias}. The noisy generalization of \eqref{eq:TSODE_SE3_GyroBias} will now be derived. The Langevin eqn.~\eqref{eq:LangevinGyroBias} may be written as
\begin{align*}
\dot{A} = \M{\se{3}}{\omega_m - \beta - \eta , - [\omega_m - \eta]_{\times} \beta + \zeta } A.
\end{align*}
The new additional term on the RHS of \eqref{eq:TangentSpaceODE_Nonlinear_SE3_GyroBias} is
\begin{align*}
e^{-\M{\se{3}}{\xi}} \left( \begin{array}{cc}
- [ \eta ]_{\times} e^{[\delta]_{\times}} e^{t [\tilde{\omega} - \wh{\beta}_0]_{\times}} R_0 & \zeta \\
0 & 0
\end{array} \right) \mu^{-1} = \left( \begin{array}{cc}
- e^{-[\delta]_{\times}} [ \eta ]_{\times} e^{[\delta]_{\times}} & e^{-[\delta]_{\times}} [ \eta ]_{\times} e^{[\delta]_{\times}} \wh{\beta}_0 + e^{-[\delta]_{\times}} \zeta \\
0 & 0
\end{array} \right) . 
\end{align*}
Upon vectorizing the previous equation and appropriately augmenting the RHS of \eqref{eq:TangentSpaceODE_Nonlinear_SE3_GyroBias2} with the result yields the desired Langevin equation on the Lie algebra:
\begin{align}\label{eq:TSSDE_SE3_GyroBias}
 \dot{\xi} = - \adbar_{\xi} \left( \begin{array}{c}
 \omega_m - \wh{\beta}_0 \\
 - [\omega_m]_{\times} \wh{\beta}_0
\end{array} \right) + \overline{W}_{\!\!\se{3}} (\xi) \left( \begin{array}{c}
 \wh{\beta}_0 - \tilde{\beta} (\delta , u , \wh{\beta}_0) - \eta \\
~ [\omega_m]_{\times} \left( \wh{\beta}_0  - \tilde{\beta} (\delta , u , \wh{\beta}_0) \right) + \zeta - [\tilde{\beta} (\delta , u , \wh{\beta}_0)]_{\times} \eta 
\end{array} \right) . 
\end{align}
Notice how the diffusion tensor depends on the MAP $\mu$ through $\wh{\beta}_0$, unlike \eqref{eq:TSSDE_2}. A compact formula for $N$ \eqref{def:N_SE3} that is convenient for the numerical propagation of moments through \eqref{eq:TSSDE_SE3_GyroBias} is presented in Appendix \ref{app:FormulaN}. 

If instead the DP group law on $\SO{3} \times \R^3$ is used, then the Langevin equation on the Lie algebra would read
\begin{align}\label{eq:TSSDE_DP_GyroBias}
\left\lbrace \begin{array}{l}
\dot{\delta} = [\omega_m -  \wh{\beta}_0]_{\times} \delta - \overline{W}_{\!\!\so{3}} (\delta) u - \overline{W}_{\!\!\so{3}} (\delta) \eta ,\\[3pt]
\dot{u} = \zeta .
\end{array} \right.
\end{align}
Comparing \eqref{eq:TSODE_SE3_GyroBias} with the deterministic part of \eqref{eq:TSSDE_DP_GyroBias}, the evolution of the Lie algebra coordinates associated with rotations (i.e., the $\delta$) is now nonlinear, while the evolution of the coordinates associated with the gyro bias is linear. This is a result of the DP group law making the vector field in \eqref{eq:LangevinGyroBias} group-affine when restricted to the bias states. This highlights a key trade-off between working with the $\SE{3}$ and DP group law on $\SO{3} \times \R^3$ for filtering. In the following, a performance comparison of an attitude filter is made using these two different choices of group laws. A dramatic improvement when working with the $\SE{3}$ group law versus the DP group law, as well as compared to the USQUE \cite{Crassidis_UKF}, is observed.

\subsection{Numerical experiment}\label{sub:experiment}

As a proof-of-principle test of attitude filtering using CGs on $\SO{3} \times \R^3$, consider an example of a satellite undergoing purely Keplerian motion with inclination $35$ [deg] and radius $6775.19$ [km]~\cite{Crassidis_UKF}. The spacecraft keeps its body $x$-axis (roll) along the velocity vector, its body $z$-axis (yaw) along the nadir, and the body $y$-axis (pitch) along the negative orbit normal. With this geometry, the orbit period is exactly $92.5$ [min] = $5550$ [s] and the body frame spacecraft angular velocity is $\omega_{\rm{orbit}} = (0, -2\pi/5550 ,0)$ [rad/sec]. The spacecraft is equipped with a strapdown, three axis gyroscope with its axes aligned with the spacecraft's body frame. Gyro measurements are simulated in accordance with the Farrenkopf model \eqref{eq:Farrenkopf} using the procedure outlined in \cite{Crassidis_SPKF}. 

Measurements of the spacecraft's attitude are given by an unbiased triaxial magnetometer with zero-mean Gaussian distributed measurement noise with covariance $\sigma_m^2 I_3$, i.e. $z = R B + n$, where $B$ is Earth's magnetic field in an inertial reference frame (to be defined), $R$ is given by is the rotation from the inertial frame to the spacecraft's body frame, and $n \sim \cal{N}^3 (0 , \sigma_m^2 I_3)$. The Earth's magnetic field along the circular orbit is modeled as a tilted dipole according to $B = \left(25.54~ [\mu\mbox{T}\right])\times \left(3 \langle m (t) , r_{\rm{unit}} \rangle r_{\rm{unit}} - m (t) \right)$, in which $r_{\rm{unit}} = r_{\rm{orbit}} / \| r_{\rm{orbit}} \|$ is a unit vector from the center of the Earth to the spacecraft and the Earth dipole is $m(t) = (\sin{(\theta_E)} \sin{(\alpha_E t)}, \sin{(\theta_E)} \cos{(\alpha_E t)}, \cos{(\theta_E)})$, where $\theta_E = 168.6$ [deg] and $\alpha_E = 4.178 \times 10^{-3}$ [deg/s].

The conditions of the spacecraft at the first measurement are that it is crossing the ascending node, and so the angle between the body $x$ axis and the equatorial plane is the orbit inclination of $\theta_i=$35$^{\circ}$. To clarify, let us take the inertial ($XYZ$) reference frame such that the ascending node lies on the negative $Y$ axis. The unit vector pointing to the orbit is then defined as $r_{\rm{orbit}} = (\cos{(\theta_i)} \sin{( \| \omega_{\rm{orbit}} \| t )} , -\cos{( \| \omega_{\rm{orbit}} \| t )} , \sin{(\theta_i)} \sin{( \| \omega_{\rm{orbit}} \| t )})$, and the velocity vector points along $v_{\rm{orbit}} = \| \omega_{\rm{orbit}} \| (\cos{(\theta_i)} \cos{( \| \omega_{\rm{orbit}} \| t )} , \sin{( \| \omega_{\rm{orbit}} \| t )} , \sin{(\theta_i)} \cos{( \| \omega_{\rm{orbit}} \| t )})$. The true initial attitude quaternion is $q_{\rm{true}} (0) = ( -0.6744 , -0.2126 , -0.2126 , 0.6744 )$.

Three attitude filtering approaches are analyzed for the scenario outlined above. Two of the filters are TSFs, one using the $\SE{3}$ group law and the other the DP group law on the state space $\SO{3} \times \R^3$. These filters are dubbed TSF-$\SE{3}$ and TSF-DP, respectively. Both filters follow the recursive algorithm summarized in Section \ref{sub:summary}, with the exception that the SDE used for propagation of the tangent space variables is \eqref{eq:TSSDE_SE3_GyroBias} for the $\SE{3}$ case, and \eqref{eq:TSSDE_DP_GyroBias} for the DP case. For both the TSF-$\SE{3}$ and TSF-DP, a fourth order Runge-Kutta method is used to integrate the sigma point ODEs (see Section \ref{sub:propagation}). The third filter is the USQUE described in \cite{Crassidis_UKF}. 

The primary metric used to compare filter performance is the statistical consistency as measured by the chi-squared value. For the CG-based filters, this is the quantity $\chi_G^2$ defined in \eqref{def:chisquareG}. For the USQUE, the estimated state error covariance for the attitude state is over (generalized) Rodrigues parameters. Hence, to compute a chi-squared statistic for the USQUE, first compute the error quaternion $q_{\rm{true}} q_{\rm{est}}^{-1}$ and then map this error quaternion to its associated Rodrigues parameter. This gives a $3$ component error vector whose associated covariance is being computed by the USQUE. Together with the error in the estimated gyro bias, a chi-squared statistic similar to that of $\chi_G^2$ may be computed. 

Table \ref{tbl:simParams} gives a complete list of the simulation parameters used. The initial state covariance in Table \ref{tbl:simParams} is for the TSF-$\SE{3}$ filter. For the TSF-DP and USQUE, this initial state covariance is transformed into an equivalent covariance using the UT. The initial attitude was sampled from a CG with MAP given by the true initial attitude and tangent space covariance given by the initial attitude covariance in Table \ref{tbl:simParams}. The initial CG covariance matrix (attitude and bias) is chosen so that the initial statistical consistency value for all filters is approximately $1$. All additional parameters for the USQUE were specifically chosen to reflect the analysis in \cite{Crassidis_UKF}.  

Figure \ref{fig:ChiSq} compares root-mean-square (RMS) of the $\chi^2$ metrics for statistical consistency between the three filters for $200$ Monte Carlo (MC) runs with randomly generated magnetometer measurements, gyro measurements, and initial attitude estimates. Smaller values indicate that the filter contains errors more faithfully, with an optimum value having a time average around $1$. The TSF-$\SE{3}$ clearly maintains close proximity to the expected value of $1$, while the TSF-DP and USQUE show dramatic overconfidence in the first hour of the simulation and appear to never quite converge to the theoretically expected value by the end of the time span. The TSF-$\SE{3}$ $\chi_G^2$ always maintains a value close $1$, which points to a far more robust representation of attitude uncertainty for large initial errors. Rapid convergence and robustness is desirable for essentially all applications.

Figure \ref{fig:RPYerr} compares the RMS of the roll, pitch, and yaw errors and their associated $3\sigma$ values for each filter. Only the first $1/2$ hour of the simulation is shown in order to highlight the differences in the errors. One may observe a clear difference in the errors between the three filters, while the $3\sigma$ values appear to be roughly the same. The USQUE shows overconfidence in each roll, pitch, and yaw direction, which is consistent with Figure \ref{fig:ChiSq}. The TSF-DP has slightly worse error in each direction compared to the TSF-$\SE{3}$, which has the lowest overall error and is consistent with the $3\sigma$ bounds (c.f.~Figure \ref{fig:ChiSq}). Even though the TSF-DP roll, pitch, and yaw errors appear to be contained within their $3\sigma$ bounds, the $\chi_G^2$ consistency metric for the TSF-DP in Figure \ref{fig:ChiSq} indicates that there are important cross correlations that are not being faithfully captured by this filter. 

Figure \ref{fig:BiasErr} compares the total estimated gyro bias error and ($3$ times) the square root of the largest eigenvalue of its associated $3 \times 3$ covariance matrix for each filter. The story here is similar to that of Figures \ref{fig:ChiSq} and \ref{fig:RPYerr}. The USQUE total estimated gyro bias error is not contained within the $3\sigma$ bound in the first hour of the simulation. The TSF-$\SE{3}$ exhibits containment throughout, which is again consistent with Figure \ref{fig:ChiSq}. Overall, the TSF-$\SE{3}$ is exhibiting near optimal performance in estimating both attitude and gyro bias, as well as the cross-correlations between the two. As a final note, all three filters exhibit similar computational speed. The primary factor contributing to runtime in the TSF-based filters appears to be the propagation step based on the CTUT. Demanding a high degree of accuracy when integrating the sigma point ODEs will slow down the filter runtime. If this is a bottleneck for real-time applications, one may restore to more crude approaches when propagating the mean and covariance through the appropriate SDEs.

\begin{table}[h]
\caption{\label{tbl:simParams} Simulation Parameters}
\begin{ruledtabular}
\begin{tabular}{ccccccc}
	{Simulation length} & {$4$ [Hr]} \\\hline
    {Gyro sampling rate} & {10 [Hz]} \\
    {Gyro ARW} & {$Q_{\eta} = (3.1623 \times 10^{-7})^2 I_3$ [rad/s$^{1/2}$]}  \\
    {Gyro RRW}  & {$Q_{\zeta} = (3.1623 \times 10^{-10})^2 I_3$ [rad/s$^{3/2}$]} \\
    {True Initial Gyro Bias} & {$(20 , ~ 20 , ~ 20)$ [deg/hr]} \\\hline
    {Magnetometer sampling rate} & {1 [Hz]} \\
    {Magnetometer measurement noise $1 \sigma$} & {50 [nT]} \\\hline
    {Initial Attitude Covariance} & {$(10\pi/180)^2 I_3$} \\
    {Initial Bias Covariance} & {$(20\pi/(180\times 3600))^2 I_3$} \\
    {Initial Attitude Estimate} & {Randomly Sampled} \\   
    {Initial Gyro Bias Estimate} & {$(0 , ~ 0 , ~ 0)$ [deg/hr]} \\\hline
    {Whitening Tolerance} & {$10^{-15}$} \\\hline 
    {UT $\lambda$} & 0 \\ 
\end{tabular}
\end{ruledtabular}
\end{table}

\clearpage

\begin{figure}[h!]
\begin{center}
    \includegraphics[width=\textwidth]{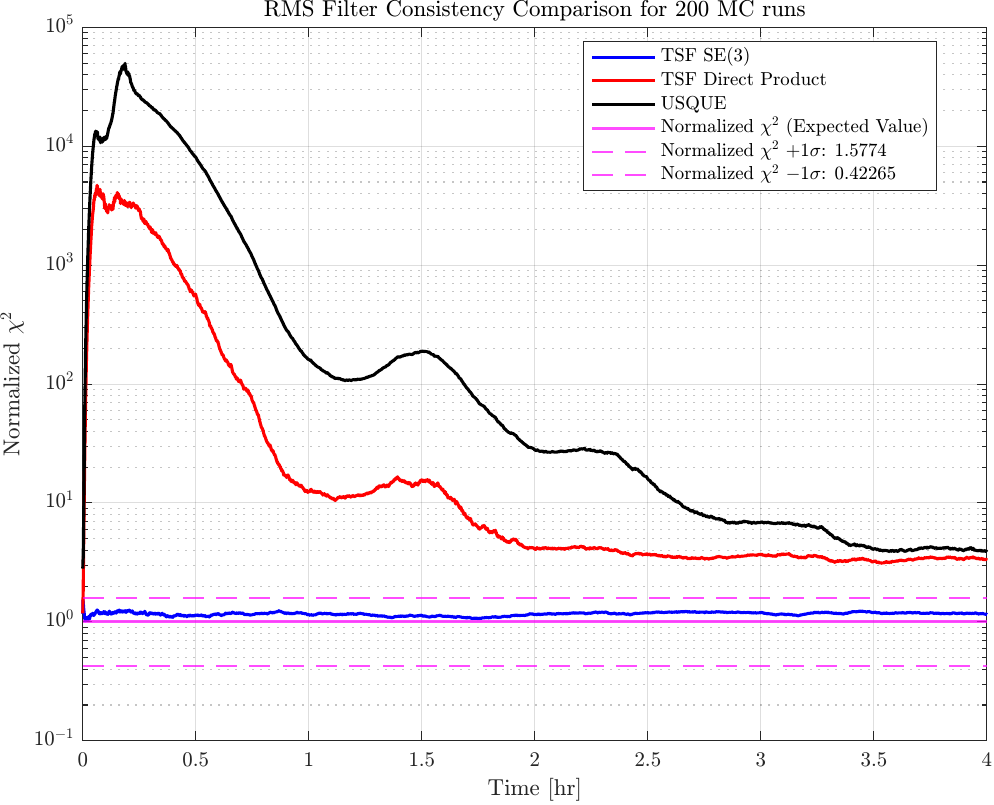} 
    \caption{The RMS over $200$ MC runs of the (normalized) $\chi^2_G$ statistical consistency metric \eqref{def:chisquareG} for the TSF-$\SE{3}$ (blue) and TSF-DP (red), along with a similar statistical consistency metric for the USQUE (black). The magenta lines represent the theoretically expected statistical consistency value for all filters, with the solid line between $1$ (the expected value) and the dashed lines representing the $\pm 1 \sigma$ around this expected value.}\label{fig:ChiSq}
    \end{center}
\end{figure}

\begin{figure}[h!]
\begin{center}
	\includegraphics[width=\textwidth]{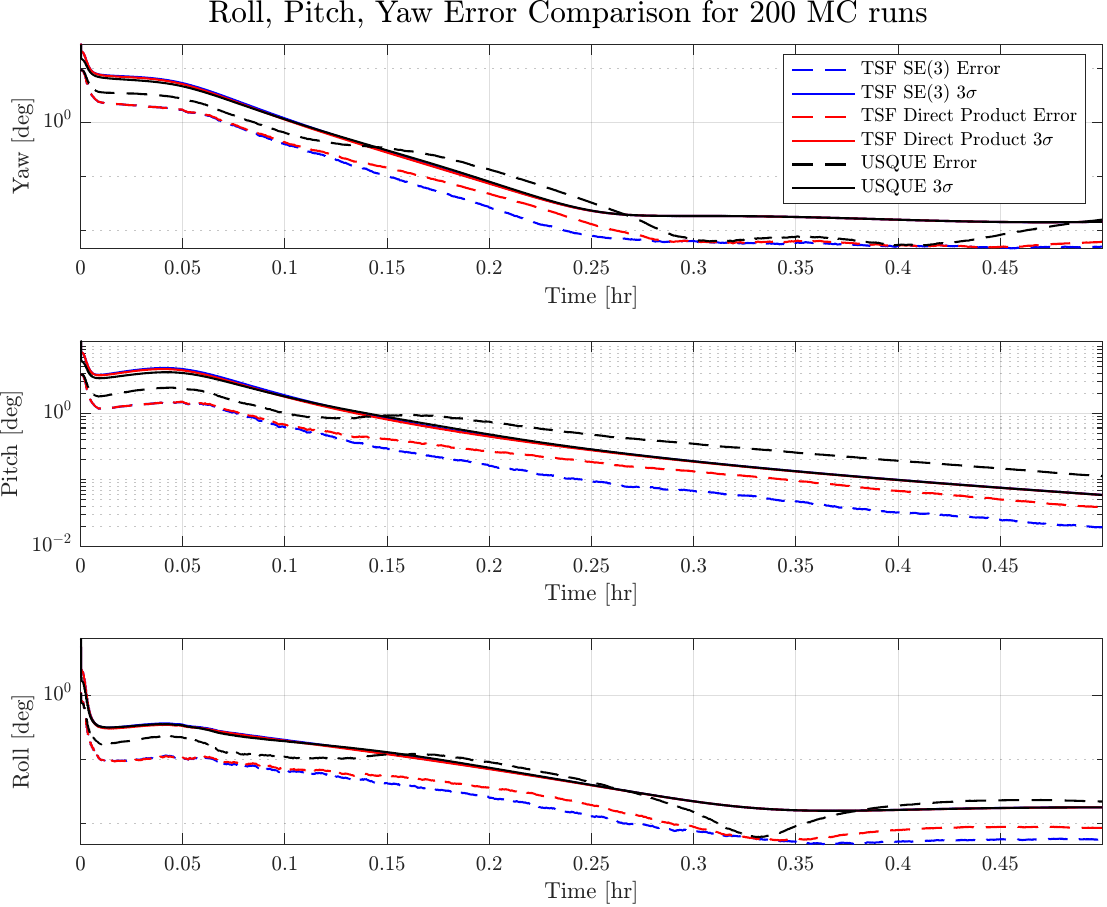} 
	\caption{The RMS over $200$ MC runs of the roll, pitch, and yaw errors and $3 \sigma$ bounds for the TSF-$\SE{3}$ (blue), TSF-DP (red), and USQUE (black). The $x$-axis only ranges over the first $1/2$ hour of the simulation in order to highlight the differences in the errors.}\label{fig:RPYerr}
\end{center}
\end{figure}

\begin{figure}[h!]
\begin{center}
	\includegraphics[width=\textwidth]{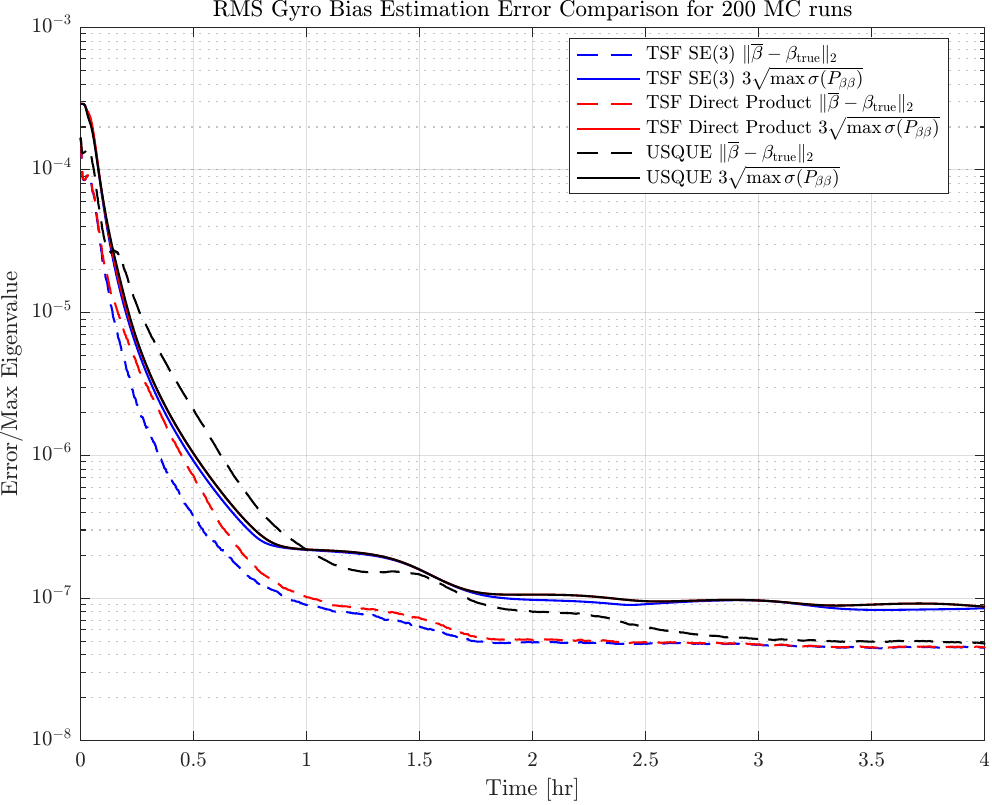} 
	\caption{The RMS over $200$ MC runs of the total gyro bias error and ($3$ times the square root of) the largest eigenvalue of the associated error covariance matrix for the TSF-$\SE{3}$ (blue), TSF-DP (red), and USQUE (black).}\label{fig:BiasErr}
\end{center}
\end{figure}

\clearpage

\section*{References}

\begin{thebibliography}{10}
\newcommand{\enquote}[1]{``#1''}
\expandafter\ifx\csname urlstyle\endcsname\relax
  \providecommand{\doi}[1]{doi:\discretionary{}{}{}#1}\else
  \providecommand{\doi}{doi:\discretionary{}{}{}\begingroup
  \urlstyle{rm}\Url}\fi

\bibitem{kalman1960new}
Kalman, R.~E., \enquote{A New Approach to Linear Filtering and Prediction
  Problems,} \emph{Journal of Basic Engineering}, Vol.~82, No.~1, 1960, pp.
  35--45, \doi{10.1115/1.3662552}.

\bibitem{kalman1961new}
Kalman, R.~E. and Bucy, R.~S., \enquote{New Results in Linear Filtering and
  Prediction Theory,} \emph{Journal of Basic Engineering}, Vol.~83, No.~1,
  1961, pp. 95--108, \doi{10.1115/1.3658902}.

\bibitem{grewal2007global}
Grewal, M.~S., Weill, L.~R., and Andrews, A.~P., \emph{Global positioning
  systems, inertial navigation, and integration}, John Wiley \& Sons, 2007,
  \doi{10.1002/0470099720}.

\bibitem{bar2004estimation}
Bar-Shalom, Y., Li, X.~R., and Kirubarajan, T., \emph{Estimation with
  applications to tracking and navigation: theory algorithms and software},
  John Wiley \& Sons, 2004, \doi{10.1002/0471221279}.

\bibitem{ristic2004beyond}
Ristic, B., Arulampalam, S., and Gordon, N., \emph{Beyond the Kalman filter:
  Particle filters for tracking applications}, Vol. 685, Artech house Boston,
  2004.

\bibitem{julier1997new}
Julier, S.~J. and Uhlmann, J.~K., \enquote{{New extension of the Kalman filter
  to nonlinear systems},} in Kadar, I., ed., \enquote{Signal Processing, Sensor
  Fusion, and Target Recognition VI,} International Society for Optics and
  Photonics, SPIE, Vol. 3068, 1997, pp. 182 -- 193,
  \doi{10.1117/12.280797}.

\bibitem{benevs1981exact}
Bene{\v{s}}, V.~E., \enquote{Exact finite-dimensional filters for certain
  diffusions with nonlinear drift,} \emph{Stochastics}, Vol.~5, No. 1-2, 1981,
  pp. 65--92, \doi{10.1080/17442508108833174}.

\bibitem{daum2005nonlinear}
Daum, F., \enquote{Nonlinear filters: beyond the Kalman filter,} \emph{IEEE
  Aerospace and Electronic Systems Magazine}, Vol.~20, No.~8, 2005, pp. 57--69,
  \doi{10.1109/MAES.2005.1499276}.

\bibitem{gordon1993novel}
Gordon, N., Salmond, D., and Smith, A., \enquote{Novel approach to
  nonlinear/non-Gaussian Bayesian state estimation,} \emph{IEE Proceedings F
  (Radar and Signal Processing)}, Vol. 140, 1993, pp. 107--113,
  \doi{10.1049/ip-f-2.1993.0015}.

\bibitem{daum2003curse}
Daum, F. and Huang, J., \enquote{Curse of dimensionality and particle filters,}
  in \enquote{2003 IEEE Aerospace Conference Proceedings (Cat. No.03TH8652),}
  Vol.~4, 2003, \doi{10.1109/AERO.2003.1235126}.

\bibitem{barrau2019linear}
Barrau, A. and Bonnabel, S., \enquote{Linear observed systems on groups,}
  \emph{Systems \& Control Letters}, Vol. 129, 2019, pp. 36--42,
  \doi{https://doi.org/10.1016/j.sysconle.2019.05.005}.

\bibitem{sarkka2007unscented}
Sarkka, S., \enquote{On Unscented Kalman Filtering for State Estimation of
  Continuous-Time Nonlinear Systems,} \emph{IEEE Transactions on Automatic
  Control}, Vol.~52, No.~9, 2007, pp. 1631--1641,
  \doi{10.1109/TAC.2007.904453}.

\bibitem{barrau2022geometry}
Barrau, A. and Bonnabel, S., \enquote{The Geometry of Navigation Problems,}
  \emph{IEEE Transactions on Automatic Control}, Vol.~68, No.~2, 2023, pp.
  689--704, \doi{10.1109/TAC.2022.3144328}.

\bibitem{Crassidis_UKF}
Crassidis, J.~L. and Markley, F.~L., \enquote{Unscented Filtering for
  Spacecraft Attitude Estimation,} \emph{Journal of Guidance, Control, and
  Dynamics}, Vol.~26, No.~4, 2003, pp. 536--542, \doi{10.2514/2.5102}.

\bibitem{shuster1993survey}
{Shuster}, M.~D., \enquote{{Survey of attitude representations},} \emph{Journal
  of the Astronautical Sciences}, Vol.~41, No.~4, 1993, pp. 439--517.

\bibitem{farrenkopf1978analytic}
Farrenkopf, R., \enquote{Analytic Steady-State Accuracy Solutions for Two
  Common Spacecraft Attitude Estimators,} \emph{Journal of Guidance and
  Control}, Vol.~1, No.~4, 1978, pp. 282--284, \doi{10.2514/3.55779}.

\bibitem{Crassidis_Extension}
Dianetti, A.~D. and Crassidis, J.~L., \emph{Extension of Farrenkopf
  Steady-State Solutions with Estimated Angular Rate},
  \doi{10.2514/6.2018-2095}.

\bibitem{hall2013lie}
Hall, B.~C., \emph{Lie groups, Lie algebras, and representations}, Springer,
  2013, \doi{10.1007/978-3-319-13467-3}.

\bibitem{muller2021review}
M{\"u}ller, A., \enquote{Review of the exponential and Cayley map on $\SE{3}$
  as relevant for Lie group integration of the generalized Poisson equation and
  flexible multibody systems,} \emph{Proceedings of the Royal Society A:
  Mathematical, Physical and Engineering Sciences}, Vol. 477, No. 2253, 2021,
  p. 20210303, \doi{10.1098/rspa.2021.0303}.

\bibitem{wolfe2011bayesian}
Wolfe, K.~C., Mashner, M., and Chirikjian, G.~S., \enquote{Bayesian fusion on
  Lie groups,} \emph{Journal of Algebraic Statistics}, Vol.~2, No.~1,
  \doi{10.18409/JAS.V2I1.11}.

\bibitem{barrau2014intrinsic}
Barrau, A. and Bonnabel, S., \enquote{Intrinsic Filtering on Lie Groups With
  Applications to Attitude Estimation,} \emph{IEEE Transactions on Automatic
  Control}, Vol.~60, No.~2, 2015, pp. 436--449,
  \doi{10.1109/TAC.2014.2342911}.

\bibitem{brossard2017unscented}
Brossard, M., Bonnabel, S., and Condomines, J.-P., \enquote{Unscented Kalman
  filtering on Lie groups,} in \enquote{2017 IEEE/RSJ International Conference
  on Intelligent Robots and Systems (IROS),} , 2017, pp. 2485--2491,
  \doi{10.1109/IROS.2017.8206066}.

\bibitem{AttitudeSurvey07}
Crassidis, J.~L., Markley, F.~L., and Cheng, Y., \enquote{Survey of Nonlinear
  Attitude Estimation Methods,} \emph{Journal of Guidance, Control, and
  Dynamics}, Vol.~30, No.~1, 2007, pp. 12--28, \doi{10.2514/1.22452}.

\bibitem{WangLee2021}
Wang, W. and Lee, T., \enquote{Bingham-Gaussian Distribution on
  $\mathbb{S}^{3}\times \mathbb{R}^{n}$ for Unscented Attitude Estimation,} in
  \enquote{2021 IEEE International Conference on Multisensor Fusion and
  Integration for Intelligent Systems (MFI),} , 2021, pp. 1--7,
  \doi{10.1109/MFI52462.2021.9591164}.

\bibitem{barfoot2014associating}
Barfoot, T.~D. and Furgale, P.~T., \enquote{Associating Uncertainty With
  Three-Dimensional Poses for Use in Estimation Problems,} \emph{IEEE
  Transactions on Robotics}, Vol.~30, No.~3, 2014, pp. 679--693,
  \doi{10.1109/TRO.2014.2298059}.

\bibitem{UnscentedKF04}
Julier, S. and Uhlmann, J., \enquote{Unscented filtering and nonlinear
  estimation,} \emph{Proceedings of the IEEE}, Vol.~92, No.~3, 2004, pp.
  401--422, \doi{10.1109/JPROC.2003.823141}.

\bibitem{sjoberg2021lie}
Sjøberg, A.~M. and Egeland, O., \enquote{Lie Algebraic Unscented Kalman Filter
  for Pose Estimation,} \emph{IEEE Transactions on Automatic Control}, Vol.~67,
  No.~8, 2022, pp. 4300--4307, \doi{10.1109/TAC.2021.3121247}.

\bibitem{ye2023uncertainty}
Ye, J., Jayaraman, A.~S., and Chirikjian, G.~S., \enquote{Uncertainty
  Propagation on Unimodular Matrix Lie Groups,} \emph{IEEE Transactions on
  Automatic Control}, pp. 1--16, \doi{10.1109/TAC.2024.3486652}.

\bibitem{ye2024uncertainty}
Ye, J. and Chirikjian, G.~S., \enquote{Uncertainty Propagation and Bayesian
  Fusion on Unimodular Lie Groups from a Parametric Perspective,} \emph{arXiv
  preprint arXiv:2401.03425}.

\bibitem{Markley_SO3}
Markley, F.~L., \enquote{Attitude filtering on $\rm{SO}(3)$,} \emph{The Journal
  of the Astronautical Sciences}, Vol.~54, No.~3, 2006, pp. 391--413,
  \doi{10.1007/BF03256497}.

\bibitem{do1992riemannian}
Do~Carmo, M.~P. and Flaherty~Francis, J., \emph{Riemannian geometry}, Vol.~6,
  Springer, 1992.

\bibitem{rosenberg1997laplacian}
Rosenberg, S., \emph{The Laplacian on a Riemannian Manifold: An Introduction to
  Analysis on Manifolds}, London Mathematical Society Student Texts, Cambridge
  University Press, 1997, \doi{10.1017/CBO9780511623783}.

\bibitem{berger2003panoramic}
Berger, M., \emph{A panoramic view of Riemannian geometry}, Springer, 2003,
  \doi{10.1007/978-3-642-18245-7}.

\bibitem{faraut2008analysis}
Faraut, J., \emph{Analysis on Lie Groups: An Introduction}, Cambridge Studies
  in Advanced Mathematics, Cambridge University Press, 2008,
  \doi{10.1017/CBO9780511755170}.

\bibitem{Schulman68}
Schulman, L., \enquote{A Path Integral for Spin,} \emph{Phys. Rev.}, Vol. 176,
  1968, pp. 1558--1569, \doi{10.1103/PhysRev.176.1558}.

\bibitem{van1978computing}
Van~Loan, C., \enquote{Computing integrals involving the matrix exponential,}
  \emph{IEEE Transactions on Automatic Control}, Vol.~23, No.~3, 1978, pp.
  395--404, \doi{10.1109/TAC.1978.1101743}.

\bibitem{barrau2016invariant}
Barrau, A. and Bonnabel, S., \enquote{The Invariant Extended Kalman Filter as a
  Stable Observer,} \emph{IEEE Transactions on Automatic Control}, Vol.~62,
  No.~4, 2017, pp. 1797--1812, \doi{10.1109/TAC.2016.2594085}.

\bibitem{Crassidis_SPKF}
Crassidis, J., \enquote{Sigma-point Kalman filtering for integrated GPS and
  inertial navigation,} \emph{IEEE Transactions on Aerospace and Electronic
  Systems}, Vol.~42, No.~2, 2006, pp. 750--756,
  \doi{10.1109/TAES.2006.1642588}.

\end{thebibliography}

\appendix

\section*{Formula for $N$}\label{app:FormulaN}

This appendix provides a formula for $N$ in \eqref{def:N_SE3} that is suitable for numerical implementation when propagating moments through the SDE \eqref{eq:TSSDE_SE3_GyroBias}. This formula is based on the following integral identity for the matrix exponential.
\begin{lem}
Let $A \in \R^{n \times n}$ be a matrix which commutes with its Moore-Penrose pseudo-inverse, $A^{\#}$ (e.g., a normal matrix). Then, 
\begin{align}\label{eq:MatrixExpInt_1}
\int_0^1 e^{-s A} \dd s = \proj_{\ker{A}} + (I_n - e^{-A}) A^{\#} ,
\end{align}
and
\begin{align}\label{eq:MatrixExpInt_2}
\int_0^1 s e^{-s A} \dd s = \frac{1}{2} \proj_{\ker{A}} + ( I_n - (I_n + A) e^{-A} ) (A^{\#})^2 .
\end{align}
\end{lem}
\textit{Proof.} In general $A^{\#}$ satisfies $A^{\#} A = I_n - \proj_{\ker{A}}$. From the assumption that $[A , A^{\#}] = 0$, this implies $A A^{\#} =  I_n - \proj_{\ker{A}}$. In other words, when $[A , A^{\#}] = 0$, then $A^{\#}$ is both the left and right inverse of $A$ when restricted to the orthogonal complement of $\ker{A}$. Therefore, 
\begin{align*}
(e^A - I) A^{\#} = \left( I + \frac{1}{2} A + \frac{1}{3!} A^2 + \cdots \right) (I - \proj_{\ker{A}}) = \left( \int_0^1 e^{t A}\dd t \right) (I - \proj_{\ker{A}}) .
\end{align*}
Hence,
\begin{align*}
\int_0^1 e^{s A}\dd s = (e^A - I) A^{\#} + \left( \int_0^1 e^{t A} \dd t \right) \proj_{\ker{A}} . 
\end{align*}
The final step is to note $\left( \int_0^1 e^{t A} \dd t \right) \proj_{\ker{A}} = \proj_{\ker{A}}$. Replacing $A$ with $-A$ results in \eqref{eq:MatrixExpInt_1}. A similar argument proves \eqref{eq:MatrixExpInt_2}. $\square$

Formula \eqref{eq:MatrixExpInt_1} does not directly apply with $A$ replaced by \eqref{eq:adbarSE3} because this matrix doesn't commute with its Moore-Penrose pseudo-inverse. However, for $\delta \in \R^3$, the Moore-Penrose pseudo-inverse of $[\delta]_{\times}$ is $- [\delta]_{\times} / \| \delta \|^2$. Together with \eqref{eq:MatrixExpInt_1}, this implies
\begin{align}\label{eq:JacobianSO3_Explicit}
\overline{J}_{\!\so{3}} (\delta) = \int_0^1 e^{-s [\delta]_{\times}} \dd s = \frac{\delta \delta^T}{\| \delta \|^2} - \frac{1}{\|\delta\|^2} ( I_3  - e^{-[\delta]_{\times}} ) [\delta]_{\times} . 
\end{align} 
Therefore,
\begin{align*}
N( \delta , u) & = \int_0^1 e^{s [\delta]_{\times}} [ \int_0^s e^{-r [\delta]_{\times}} u \dd r ]_{\times} \dd s \\
& = \int_0^1 e^{s [\delta]_{\times}} [ \left( s \frac{\delta \delta^T}{\| \delta \|^2} - \frac{1}{\|\delta\|^2} ( I_3  - e^{-s[\delta]_{\times}} ) [\delta]_{\times} \right) u ]_{\times} \dd s \\
& = \frac{1}{\| \delta \|^2} \left( \int_0^1 s e^{s [\delta]_{\times}} \dd s \right) [\delta \delta^T u]_{\times} - \frac{1}{\|\delta\|^2} \int_0^1 e^{-s [\delta]_{\times}} [ ( I_3 - e^{-s[\delta]_{\times}}) [\delta]_{\times} u ]_{\times} \dd s . \numberthis \label{eq:GyroBiasSE3_derivation16}
\end{align*}
From \eqref{eq:MatrixExpInt_2}, 
\begin{align}\label{eq:GyroBiasSE3_derivation17}
\int_0^1 s e^{s [\delta]_{\times}} \dd s = \frac{1}{2} \frac{\delta \delta^T}{\| \delta \|^2} - \frac{1}{\|\delta\|^4} ( I_3 - (I_3 - [\delta]_{\times}) e^{[\delta]_{\times}} ) [\delta]_{\times}^2 . 
\end{align}
Plugging \eqref{eq:GyroBiasSE3_derivation17} into the first term on the RHS of \eqref{eq:GyroBiasSE3_derivation16} and simplify, one finds
\begin{align*}
\frac{1}{\| \delta \|^2} \left( \int_0^1 s e^{s [\delta]_{\times}} \dd s \right) [\delta \delta^T u]_{\times} = - \frac{\langle \delta , u \rangle}{\|\delta\|^4} ( I_3 - (I_3 - [\delta]_{\times}) e^{[\delta]_{\times}} ) [\delta]_{\times} .
\end{align*}
For the next term on the RHS of \eqref{eq:GyroBiasSE3_derivation16}, first observe that
\begin{align}\label{eq:GyroBiasSE3_derivation18}
e^{s [\delta]_{\times}} [ ( I_3 - e^{-s[\delta]_{\times}}) [\delta]_{\times} u ]_{\times} \dd s = \left[ e^{s [\delta]_{\times}} , [ [\delta]_{\times} u ]_{\times} \right] ,
\end{align}
where the outer most brackets in \eqref{eq:GyroBiasSE3_derivation18} represents the matrix commutator. Hence,
\begin{align*}
\int_0^1 e^{s [\delta]_{\times}} [ ( I_3 - e^{-s[\delta]_{\times}}) [\delta]_{\times} u ]_{\times} \dd s = \left[ \int_0^1 e^{s [\delta]_{\times}} \dd s , [ [\delta]_{\times} u ]_{\times} \right] .
\end{align*}
In total,
\begin{align}\label{eq:NSE3_explicit}
N (\delta , u) = \frac{1}{\|\delta\|^2} \left[  [ [\delta]_{\times} u ]_{\times} , \overline{J}_{\!\so{3}} (-\delta) \right] - \frac{\langle \delta , u \rangle}{\|\delta\|^4} ( I_3 - (I_3 - [\delta]_{\times}) e^{[\delta]_{\times}} ) [\delta]_{\times} .
\end{align}
Formula \eqref{eq:NSE3_explicit} is convenient for numerical implementation because it doesn't involve explicit trigonometric function through, e.g. the Rodrigues' rotation formula \eqref{eq:Rodrigues}, and it reuses quantities that will already be necessary to compute (e.g., $\overline{J}_{\!\so{3}} (-\delta)$). Moreover, it is easier to analytically compute derivatives of \eqref{eq:NSE3_explicit}, which is necessary to compute the drift correction in \eqref{eq:tildef_def}. 

\end{document}